%% file: main.tex
\documentclass[10pt, sigconf]{acmart}
\renewcommand\footnotetextcopyrightpermission[1]{}
\newcommand{\ignore}[1]{}

\newcommand{\ProveIt}{\textbf{Prove-It}}
\newcommand{\ProveItWebsite}{\textbf{Prove-It} website \cite{website:Prove-It}}
\newcommand{\exprtype}[1]{\textbf{#1}}

\newcommand{\infrule}[1]{\textbf{#1}}
\newcommand{\reductionrule}[1]{\textbf{#1}}
\newcommand{\subexpr}[1]{\emph{#1}}

\usepackage{amsmath,amssymb,amsfonts}
\usepackage{graphicx}
\usepackage[font=footnotesize,labelfont=bf,textfont=footnotesize]{caption}

\usepackage[size=footnotesize]{subcaption} 
\usepackage{textcomp}
\def\BibTeX{{\rm B\kern-.05em{\sc i\kern-.025em b}\kern-.08em
    T\kern-.1667em\lower.7ex\hbox{E}\kern-.125emX}}

\usepackage{booktabs}  

\usepackage{url}
\usepackage[framemethod=tikz]{mdframed}
\mdfdefinestyle{MyFrame}{%
    linecolor=black,
    outerlinewidth=2pt,
    innertopmargin=4pt,
    innerbottommargin=4pt,
    innerrightmargin=4pt,
    innerleftmargin=4pt,
        leftmargin = 4pt,
        rightmargin = 4pt
}

\usepackage{comment}

\usepackage{syllogism}
\usepackage{proof}
\usepackage{turnstile}
\usepackage{appendix}
\usepackage{enumitem}

\usepackage{pdfpages}

\begin{document}

\setlength{\abovedisplayskip}{3pt}
\setlength{\belowdisplayskip}{3pt}
\setlength{\abovedisplayshortskip}{1pt}
\setlength{\belowdisplayshortskip}{2pt}

\DeclareRobustCommand\vdashvar{\mathrel{|}\joinrel\mkern-.5mu\mathrel{-}}

\title{Prove-It: A Proof Assistant for Organizing and Verifying General Mathematical Knowledge}

\author{Wayne M. Witzel}
\affiliation{%
  \small
  \institution{Center for Computing Research, Quantum Computer Science\\Sandia National Laboratories}
  \city{Albuquerque}
  \state{NM}
}
\email{wwitzel@sandia.gov}

\author{Warren D. Craft}
\affiliation{%
  \small
  \institution{Department of Computer Science,\\The University of New Mexico}
  \city{Albuquerque}
  \state{NM}
}
\email{wdcraft@unm.edu}

\author{Robert D. Carr}
\affiliation{%
  \small
  \institution{Department of Computer Science,\\The University of New Mexico}
  \city{Albuquerque}
  \state{NM}
}
\email{bobcarr@unm.edu}

\author{Joaquín E. Madrid Larrañaga}
\affiliation{%
  \small
  \institution{Center for Computing Research, Quantum Computer Science\\Sandia National Laboratories}
  \city{Albuquerque}
  \state{NM}
}
\email{jmadri@sandia.gov}

\begin{abstract}
We introduce \ProveIt{}, a Python-based general-purpose interactive theorem-proving assistant designed with the goal of making formal theorem proving as easy and natural as informal theorem proving (with moderate training).  \ProveIt{} uses a highly-flexible Jupyter notebook-based user interface that documents interactions and proof steps using \LaTeX. We review Prove-It's highly expressive representation of expressions, judgments, theorems, and proofs; demonstrate the system by constructing a traditional proof-by-contradiction that $\sqrt{2}\notin\mathbb{Q}$; and discuss how the system avoids inconsistencies such as Russell's and Curry's paradoxes. Extensive documentation is provided in the appendices about core elements of the system. Current development and future work includes promising applications to quantum circuit manipulation and quantum algorithm verification.

\end{abstract}

\keywords{automated theorem proving, proof assistant, python, Jupyter Notebook, quantum algorithm, quantum circuit}

\maketitle
\thispagestyle{plain}
\pagestyle{plain}


\section{Introduction \& Motivation}

\ProveIt{} is an open-source Python library that has been designed as a general-purpose theorem prover to support structured proof~\cite{Lamport:2012} development to build formal proofs in a manner that mimics informal proof development.
A primary target application for the Prove-It tool is the verification and analysis of quantum circuits and software, though it is designed as a general-purpose theorem prover.
Quantum algorithm and circuit verification is a new but active area of research~\cite{Maslov:2008, Nam:2018, Ardeshir:2014_verification, Venturelli:2019, Hietala:2019, Liu:2016_theorem_prover, Burgholzer:2020_verifying_IBM_Qiskit} as small quantum computing machines have become publicly available~\cite{website:IBM_Quantum_Experience, Preskill:2018_NISQ}.  The \ProveIt{} approach is unique among the various other theorem provers~\cite{Wiedijk:2006_17_Provers} in its versatility, with a design intended to extend to any form of logical reasoning in any subject.

Flexibility has been a key design principle guiding the development of \ProveIt{}, making it particularly well-suited to handle the various abstract concepts of quantum information theory. A prover for this application must not only support basic logical reasoning about propositional connectives and quantifiers over classical data values, including numbers and sets, but also express sophisticated concepts used in quantum computing, particularly linear algebra (matrices, tensor products, and unitary transformations) and probability theory (expectation values), as well as quantification over such constructs.

Another consideration is that the theorem prover is anticipated to be used by physicists and application experts in quantum computation who may not have training in formal methods.  One of the main concerns in design decisions has thus been ease of proof construction (with sensible and predictable automation) and use of notation commonly used by physicists and mathematicians interested in scientific applications.

A guiding principle in the development of Prove-It has been to support the formalization of informal proofs (e.g., sketched in text books and journals) as a nearly direct translation.
Ideally, with enough development in a particular subject area, 
it should be possible to develop formal proofs with no more difficulty than expressing an informal proof at a high level of abstraction.  
Most existing proof checkers~\cite{Wiedijk:2007_QED_Revisited, Wiedijk:2006_17_Provers} are not likely to succeed at such proof attempts, either they do not support a sufficient level of sophisticated reasoning or gaps among various proof steps are too nontrivial to be automated.
It is hoped that a user can formalize an informal proof at the same level of detail and fill the gaps in reasoning among various steps using additional separate but small proofs (which may be re-used for other purposes, and therefore the breadth of capabilities of Prove-It will grow over time). As an exercise, as well as a demonstration of a proof of this concept, a formal proof of the accuracy and bound of distribution of a quantum phase estimation algorithm was developed on a previous incarnation of Prove-It, mimicking the informal proof given in \cite[pp 221--225]{Nielsen_Chuang:2010}. This effort, while somewhat \textit{ad hoc} and brute-force, involved 29 small proofs totalling 1750 steps by using over 275 assumptions, and was found to be rewarding since it found 3 minor mistakes in the informal proof in the book as well as managed to improve one of the bounds of the informal proof~\cite{ProveItQPE}.

Human expert knowledge is an important resource for constructing meaningful proofs.  Historically, formal methods have had a strong emphasis on automation.  In \ProveIt{}, the primary emphasis is on making the system intuitive so that we can glean the most value from human expertise.  This emphasis is still compatible with powerful automation capabilities, but it is a guiding principle that has strongly influenced the design of our system.  In many ways, \ProveIt{} is an experiment in seeing what is possible in a formal logic system that prioritizes human intuition over automation in its design.

In the early development of \ProveIt{}, we derived inspiration from the \textbf{The QED Manifesto}~\cite{Boyer:1994_QED_Manifesto} and \textbf{The QED Manifesto Revisited}~\cite{Wiedijk:2007_QED_Revisited}.  The former articulated the benefits to society that would derive from compiling a formalized computer database of all mathematical knowledge (the ``QED system'').  The latter explains why no such system exists to any great deal of satisfaction.  It cites the lack of investment into a concerted effort to bring this about as well as the fact that (existing) formal systems fail to resemble standard mathematics (and instead look like computer code).  We have made a concerted effort to address the latter issue in \ProveIt{} by employing flexible \LaTeX{} formatting of our formal expressions; while the \LaTeX{} formatting merely serves as a superficial dressing on top of the actual formal representations, real mathematical notation presents a very intuitive interface that adds a great deal of value to the system.  A suggested solution to the former issue is to create a  ‘Wikipedia for formalized mathematics’~\cite{Wiedijk:2007_QED_Revisited}.  It is our hope that our open-source software will transition into a phase of rapid expansion, in proofs and proof strategies, through contributions from an increasing number of researchers that will mutually benefit by strengthening the system.  This is an ambitious goal, but even if \ProveIt{} does not succeed at becoming the ``QED system,'' it would be worthwhile to inspire the system that does.

This paper is organized as follows.  In \S\ref{Sec:DomainsOfDiscourse}, we have an important discussion regarding ``domains of discourse.''  A distinguishing feature of the core logic of \ProveIt{} is that the ``domain of discourse'' is never presumed or enforced by a type system; instead, domain restrictions are explicitly expressed in the \emph{expressions} that form the \emph{judgments} that form the \emph{proofs}.
\S\ref{Sec:Core} explores the ``core derivation system'' of \ProveIt{} which refers to the information needed to understand and independently verify generated proofs and is intended to be fairly small and stable.  \S\ref{Sec:TheoryPackageSystem} explores the theory package system, \ProveIt{}'s ever-expanding library of knowledge.
Layers of encapsulation, abstraction, and automation are constructed in this interdependent theory system; \emph{theorems} may be proven from \emph{axioms}, \emph{conjectures} may be posited and used to construct partial proofs of other conjectures, and methods may be written to instantiate
axioms/theorems/conjectures in a convenient manner.
\S\ref{Sec:Example} demonstrates a proof that the square root of $2$ is not rational, intended as an example that a formal proof construction can have a close resemblance to an informal proof.  In \S\ref{Sec:Paradoxes}, we address some questions about consistency and paradoxes.  We conclude in \S\ref{Sec:Conclusion}.  Appendices \ref{appendix_ExprTypes}-\ref{appendix_ReductionRules} provide a thorough guide to the core derivation system.  Appendix \ref{sec:Axioms} shows some of our basic theory packages and their axioms.


\section{An Important Note Regarding Domains of Discourse}
\label{Sec:DomainsOfDiscourse}
\input{DomainsOfDiscourse}


\section{Basic Logic: The Core Derivation System}
\label{Sec:Core}

\noindent\ProveIt{} has a core system that is relatively small and intended to be fairly stable, and a theory package system that is an ever-expanding library of knowledge.  The ``core derivation system'' refers to the information needed to understand and independently verify the proofs generated by \ProveIt{}.  The details of this system are provided in Appendices~\ref{Sec:ExprTypes}, \ref{Sec:Style}, \ref{sec:InferenceRules}, and \ref{sec:ReductionRules}.  In this section, we present the important, main concepts and provide useful examples.

\subsection{Expressions and Proofs as DAGS}

\ProveIt{} uses objects, using Python's object-oriented language features, to represent \textit{judgments} as they were described in \S\ref{Sec:DomainsOfDiscourse}.  A judgment object is composed of \textit{expression} objects to represent each assumption on the left of the turnstile and the one consequent expression on the right of the turnstile.  Each judgment also has an associated proof object for deriving the judgment from \textbf{\textit{axioms}}, proven \textbf{\textit{theorems}}, and/or unproven \textbf{\textit{conjectures}} that will be discussed later.

For example, in \ProveIt{}'s \texttt{absolute\_value} context, one can derive the following judgment:
\begin{align}
    \{x\in R^{\ge 0}, y\in R^{\ge 0}\} \vdash |x \cdot y| = (x \cdot y).
    \label{kt:abs_elim_example}
\end{align}
That is: assuming that $x$ and $y$ are non-negative real numbers, then the absolute value of their product is simply the product itself (without the absolute value).\footnote{In the Jupyter Notebook-based user interface, and the \ProveItWebsite{}, a user can then simply click on the turnstile $\vdash$ symbol to obtain a detailed listing of the steps in the proof of the judgment.}  This judgment is composed of three separate expressions: $x\in R^{\ge 0}$, $y\in R^{\ge 0}$, and $|x \cdot y| = (x \cdot y)$.

\subsubsection{An Expression is a DAG}
Mathematical expressions such as $|x \cdot y|= (x \cdot y)$, $x+y$, or $\forall_{x}P(x)$ are the basic building blocks of proofs and judgments.  Each expression in \ProveIt{} is represented internally by a directed acyclic graph (DAG) of expression and sub-expression objects.

\begin{figure}[ht]
\captionsetup{width = 6cm}
\centering
    \includegraphics[width = 0.25\textwidth]{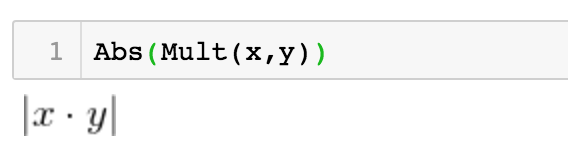}
\caption{Input code and \LaTeX-formatted output in one of Prove-It's interactive Python-based Jupyter notebooks for the expression 
$|x \cdot y|$.}
\label{fig:abs_xy_code_and_output}
\end{figure}

Consider, for example, the mathematical expression $|x \cdot y|$ that appeared as part of judgment (\ref{kt:abs_elim_example}) above. Figure~\ref{fig:abs_xy_code_and_output} shows how the input code and \LaTeX-formatted output expression would look like in one of \ProveIt{}'s interactive Python-based Jupyter notebooks.

\begin{figure*}[htb]
\captionsetup{width = 12cm}
\centering
  \begin{subfigure}{0.3\textwidth}
  \centering
  \includegraphics[width = 0.9\textwidth]{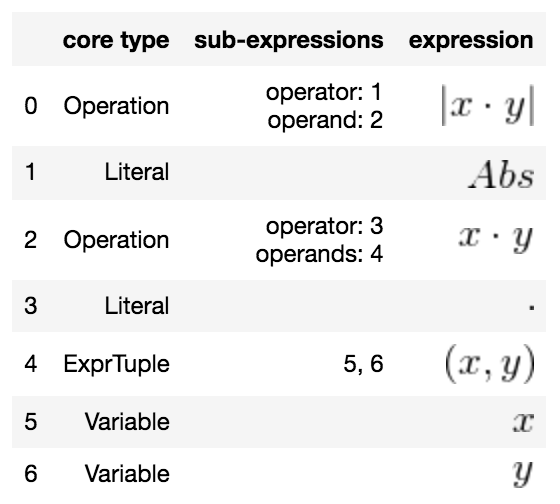}
  \caption{}
  \end{subfigure}
  \begin{subfigure}{0.22\textwidth}
  \centering
  \includegraphics[width = 0.9\textwidth]{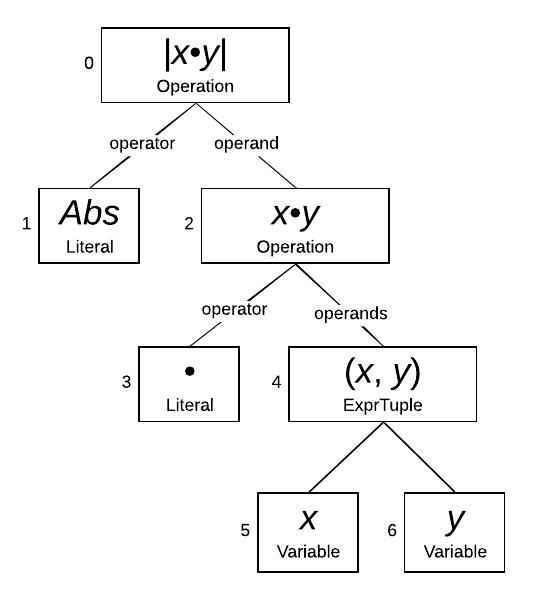}
  \caption{}
\end{subfigure}
\caption{(\textit{a}) \ProveIt{}'s table representation of the directed acyclic graph (DAG) for the expression $|xy|$. (\textit{b}) The corresponding graphical depiction of the DAG represented in part (\textit{a}). The numerical notation to the left of each boxed node indicates the corresponding line in the expression table.}
\label{fig:abs_xy_example_DAG}
\end{figure*}

In the Jupyter notebook, the user can click on the resulting \LaTeX-formatted output expression $|x \cdot y|$ to inspect the DAG construction (in tabular form) representing the expression, as shown in Figure~\ref{fig:abs_xy_example_DAG}(\textit{a}). That tabular summary corresponds to the DAG diagrammed in Figure~\ref{fig:abs_xy_example_DAG}(\textit{b}), with the numerical notations next to each node indicating the corresponding rows of the \ProveIt{} table.

\begin{figure*}[htb]
\captionsetup{width = 12cm}
\centering
  \begin{subfigure}{0.4\textwidth}
  \centering
  \includegraphics[width = 0.9\textwidth]{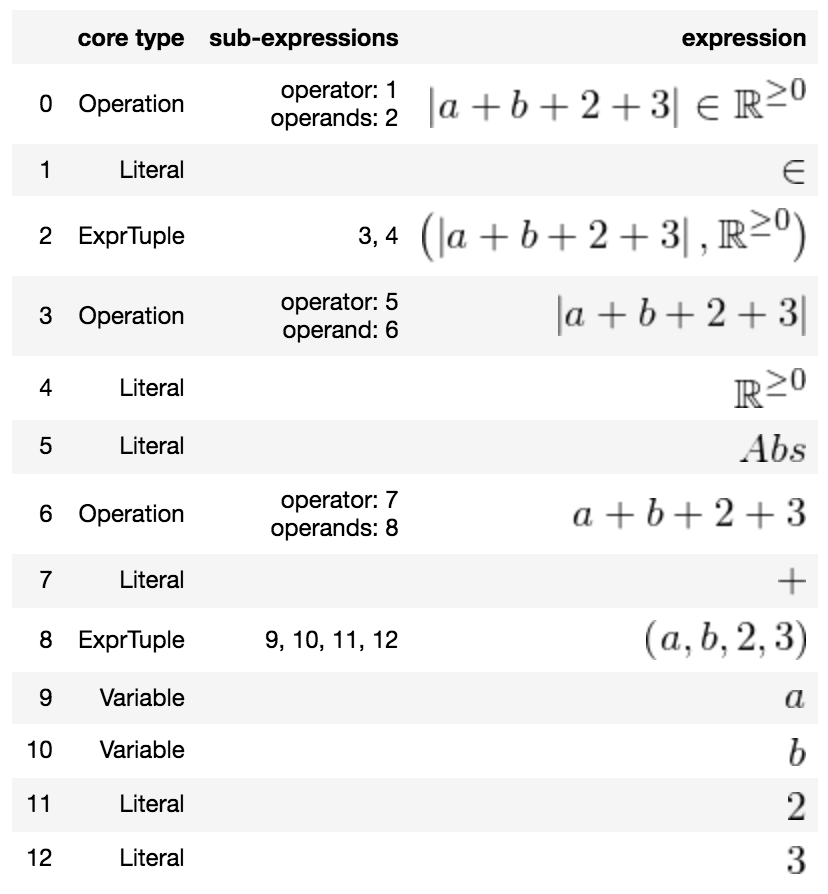}
  \caption{}
  \end{subfigure}
  \begin{subfigure}{0.32\textwidth}
  \centering
  \includegraphics[width = 0.9\textwidth]{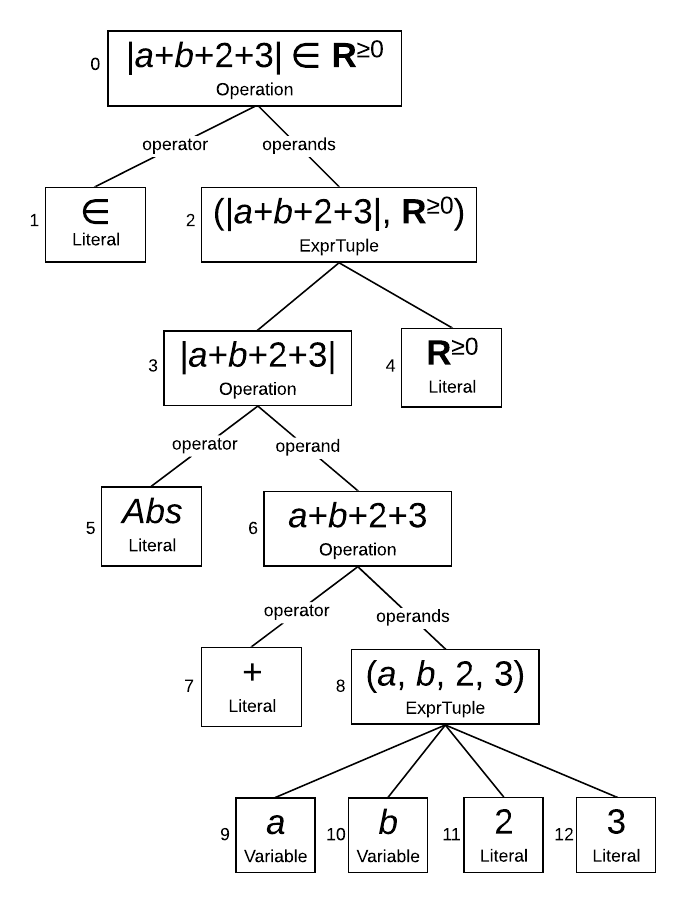}
  \caption{}
\end{subfigure}
\caption{(\textit{a}) \ProveIt{}'s table representation of the directed acyclic graph (DAG) for the expression $|a+b+2+3| \in \mathbf{R}^{\ge 0}$. (\textit{b}) A graphical depiction of the DAG shown in part (\textit{a}). The numerical notation to the left of each boxed node indicates the corresponding line in the expression table.}
\label{fig:abs_sum_ab23_example_DAG}
\end{figure*}

The DAG for the somewhat more complex expression $|a+b+2+3|\in\mathbf{R}^{\ge 0}$ is shown in Figure (\ref{fig:abs_sum_ab23_example_DAG}) in both tabular and graphical forms. Where appropriate, the DAG edges have been labeled to describe the way in which one node contributes to the formation of a node just above. For example, the Literal \texttt{+} in node \texttt{7} serves as the operator on the four operands shown in node \texttt{8} to produce the summation operation shown in node \texttt{6}. Those ``operator'' and ``operand'' notations correspond to the same notations appearing in the table version of the DAG (see row 6 of Figure~\ref{fig:abs_sum_ab23_example_DAG}(\textit{a})).

The ``core type'' columns of the Figure~\ref{fig:abs_xy_example_DAG}(\textit{a}) and Figure~\ref{fig:abs_sum_ab23_example_DAG}(\textit{a}) list one of nine primitive expression types for each node of the expression DAGs: \exprtype{Variable}, \exprtype{Literal}, \exprtype{ExprTuple}, \exprtype{Operation}, \exprtype{Conditional}, \exprtype{Lambda}, \exprtype{NamedExprs}, \exprtype{ExprRange}, and \exprtype{IndexedVar}.
These are described in Appendix~\ref{appendix_ExprTypes}.  Each of the nine primitive expression types has its own rules in the derivation system for building proofs and is therefore part of the ``core derivation system'' needed to independently verify the proofs generated by \ProveIt{}.  New expression classes may be derived from these nine (\textit{e.g.}, \texttt{Abs} and \texttt{Mult} in Figure \ref{fig:abs_xy_code_and_output} are derived from \exprtype{Operation} and provided with their own convenient methods and formatting options [see Appendix~\ref{appendix_Style}]), but the derivation rules [described in Appendices~\ref{appendix_InferenceRules} and \ref{appendix_ReductionRules}] are entirely dictated by their ``core type'' identities).  

\begin{figure*}[htb]
\captionsetup{width = 10cm}
\centering
  \includegraphics[width = 0.7\textwidth]{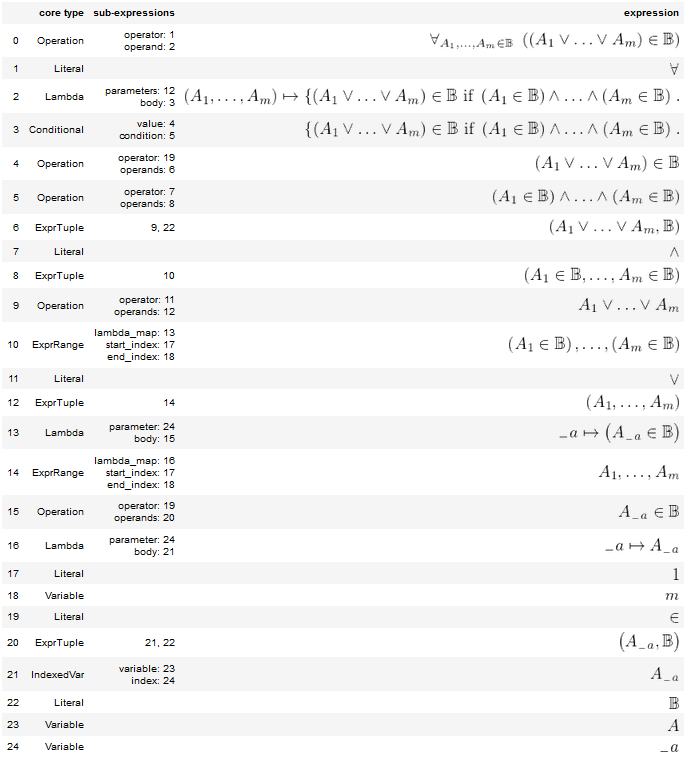}
\caption{\ProveIt{}'s table representation of the directed acyclic graph (DAG) for the expression 
$\forall_{A_{1}, \ldots, A_{m} \in \mathbb{B}}~\left(\left(A_{1} \lor \ldots \lor A_{m}\right) \in \mathbb{B}\right)$. The expression makes use of eight of \ProveIt{}'s nine possible primitive expression types.}
\label{fig:disjunction_closure_expr_DAG}
\end{figure*}

These nine primitive expression types provide for substantial expressivity.  For example,
\begin{align}
    \forall_{m \in \mathbb{N}^+}~\left[\forall_{A_{1}, \ldots, A_{m} \in \mathbb{B}}~\left(\left(A_{1} \lor \ldots \lor A_{m}\right) \in \mathbb{B}\right)\right]
    \label{expr:disjunction_closure}
\end{align}
expresses that for any positive, whole number $m$, and any $A_1, \ldots, A_m$ that are each Boolean valued, their disjunction (logical or) is a Boolean value.  Figure~\ref{fig:disjunction_closure_expr_DAG} shows the expression DAG for the sub-expression
\begin{align*}
    \forall_{A_{1}, \ldots, A_{m} \in \mathbb{B}}~\left(\left(A_{1} \lor \ldots \lor A_{m}\right) \in \mathbb{B}\right).
\end{align*}
Eight of the nine primitive expression types are used in this expression (all but \exprtype{NamedExpr}).  $m$ is a \exprtype{Variable}, \exprtype{ExprRange}s and \exprtype{IndexedVar}s are used to express $A_{1}, \ldots, A_{m}$ and
$A_1 \lor \ldots \lor A_m$. The conditional quantifier
\begin{align*}
    \forall_{A_{1}, \ldots, A_{m} \in \mathbb{B}}~\ldots
\end{align*}
is internally represented as an \exprtype{Operation} with $\forall$ as a \exprtype{Literal} operator and
\begin{align*}
    (A_1, \ldots, A_m) \mapsto \ldots~\textrm{if}~(A_1 \in \mathbb{B}) \land \ldots \land (A_m \in \mathbb{B})
\end{align*}
as a \exprtype{Lambda} operand.  Within this \exprtype{Lambda} operand is a \exprtype{Conditional} (explicitly denoted with ``if'').  $\left(A_{1}, \ldots, A_{m}\right)$ is an \exprtype{ExprTuple}.  $\mathbb{B}$ is a \exprtype{Literal}.  And so on.  While many of these primitive types are obscured by the flexible formatting, the structure that is revealed by the expression DAG with respect to primitive types is key to fully understanding \ProveIt{} derivations.

Despite the apparent complexity of the expression, it is worth noting that the interactive user \textit{construction} of such an expression is quite straightforward, with the actual \ProveIt{} notebook code looking like this:
\vspace{0.1in}
{\footnotesize
\begin{mdframed}[hidealllines=true, backgroundcolor=gray!10]
\begin{verbatim}
Forall(m,
       Forall(A_1_to_m,
              inBool(Or(A_1_to_m)),
              domain=Booleans),
       domain=NaturalsPos)
\end{verbatim}
\end{mdframed}
}
\noindent and utilizing nested \texttt{Forall} objects and an imported pre-defined expression \texttt{A\_1\_to\_m} representing the $A_1,\ldots,A_m$.

\subsubsection{A Proof is a DAG}

\begin{figure*}[htb]
\captionsetup{width = 14cm}
\centering
  \begin{subfigure}[b]{0.35\textwidth}
  \centering
  \includegraphics[width = 1.0\textwidth]{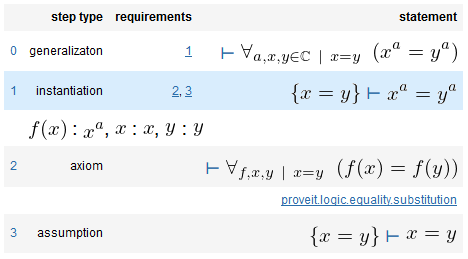}
  \caption{}
  \end{subfigure}
  \begin{subfigure}[b]{0.5\textwidth}
  \centering
  \includegraphics[width = 0.75\textwidth]{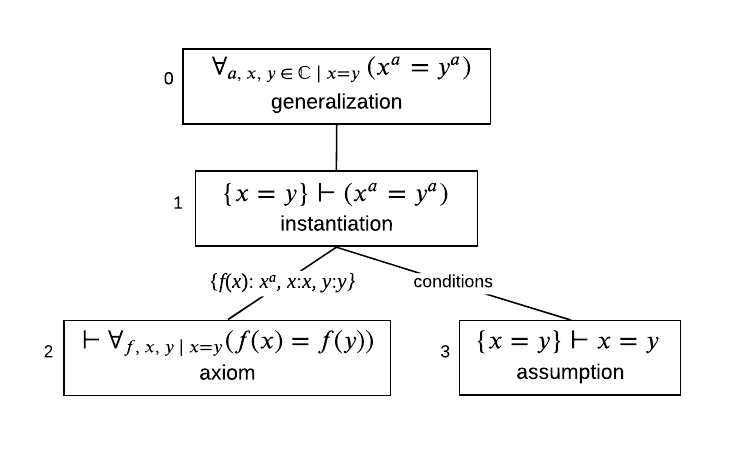}
  \caption{}
\end{subfigure}
\caption{(\textit{a}) \ProveIt's table representation of the directed acyclic graph (DAG) for the proof of the \texttt{exp\_eq} theorem $\forall_{a,x,y\in\mathbb{C}\,\rvert\, x=y}\left(x^a = y^a\right)$ (which appears in the \texttt{proveit.numbers.exponentiation} package and may be viewed online at \url{http://pyproveit.org/packages/proveit/numbers/exponentiation/_theory_nbs_/proofs/exp_eq/thm_proof.html}). (\textit{b}) A graphical depiction of the proof DAG represented in part (\textit{a}). The number to the left of each boxed node indicates the corresponding line in the \ProveIt{} proof table.}
\label{fig:exp_eq_judgment_proof_DAG}
\end{figure*}

Like an expression, a \textbf{\textit{proof}} or \textbf{\textit{proof object}} in \ProveIt{} is represented internally as a directed acyclic graph (DAG). For example, as part of Prove-It's \texttt{number/exponential} theory package, we have the following exponentiation theorem:
\begin{align}
    \vdash \forall_{a,x,y\in\mathbb{C}\,\rvert\, x=y}\left(x^a = y^a\right),
    \label{judgment:exp_eq}
\end{align}
\noindent the proof of which appears in Figure~\ref{fig:exp_eq_judgment_proof_DAG} in both the table format produced in \ProveIt{} and in a graphical representation of the underlying DAG.

Each node of a proof object DAG is a judgment. The subtree rooted at a node represents the proof of the judgment at that node. The leaf nodes are populated with assumptions, axioms, theorems, and conjectures.

Proof DAGs are derivations that are presented in a reverse order from a usual derivation.  
This reversal reflects the tree-like structure of a proof where the root is the judgment being proven.  It also offers a useful hierarchical view of a proof where you can easily focus on the important parts necessary to reach a conclusion, only digging into the details as needed.\footnote{The turnstile hyperlinks in the Jupyter Notebook-based user interface and the \ProveItWebsite{} are particular convenient for exploring proofs with this hierarchical view.}

At the root node (labeled \texttt{0}) in Figure~\ref{fig:exp_eq_judgment_proof_DAG}(\textit{b}), for example, we have the exponentiation theorem being proven, along with the ``step type'' of ``generalization'' indicating how node \texttt{0} is derived from nodes further down the tree. Contributing to the proof, we see at node \texttt{2} the general \texttt{substitution} axiom, and at node \texttt{3} the assumption that $x=y$. The edges are labeled to describe the way in which one node contributes to the derivation of the node above. For example, node \texttt{1} is derived by instantiating the substitution axiom at node \texttt{2} using the \textit{instantiation (or replacement) map} $\{f(x):x^a, x:x, y:y\}$, and using the assumption at node \texttt{3} to satisfies the $x=y$ condition required for this instantiation.

Similar to the ``core type'' identifications of an expression DAG, each node of the proof DAG is identified by a ``step type.'' There are six possibilities corresponding to six inference rules described in Appendix~\ref{appendix_InferenceRules}: \infrule{proof by assumption}, \infrule{axiom/theorem/conjecture invocation}, \infrule{modus ponens}, \infrule{deduction}, \infrule{instantiation}, and \infrule{generalization}.
Among these, \infrule{instantiation}, which is used to instantiate universally quantified variables, is the most versatile and intricate, requiring additional expression-dependent reduction rules described in Appendix~\ref{appendix_ReductionRules}.  Any derivation step represented by a node in a proof DAG can be verified using an understanding of these inference and reduction rules as well as a knowledge of the expression DAG structures of the judgements of the node and its direct children.

The `A' in DAG is very important.  A valid proof must not contain circular logic in which a judgment (apart from leaf nodes) is proven using itself directly or indirectly.  To make it easy to verify the acyclic nature of a proof DAG, requirements for a particular step will always appear after that step in the tree (\textit{i.e.}, at a node further down or higher-numbered in the DAG).  If multiple steps in the proof DAG have the same requirement, that requirement will appear after the last step that requires it. \footnote{For convenience in exploring proofs and sub-proofs (e.g., using turnstile hyperlinks), a proxy for all of the root node requirements will appear immediately following the root node, but some of these may simply reference duplicate judgments that are also required by steps further down (that is, instead of a regular ``inference rule'' step type, it will simply be marked by ``reference'' and have a single requirement for the later step with the same judgment).  This makes it easy to view, together, all of the information that pertains to the root node derivation step while also enforcing the rule that requirements of a step always appear as a later step.}

Theorem proofs will be discussed in the next section.  For a theorem proof, we must not only ensure that the proof DAG is acyclic, we also need to ensure that there is no circularity in the dependencies among the theorems.


\section{The Theory Package System}\label{sect_theory_systems}
\label{Sec:TheoryPackageSystem}

The theory package system (TPS) of \ProveIt{} is an ever-expanding library of knowledge, built upon a relatively small, stable foundational core.  The TPS is a hierarchical collection of theory packages, each consisting of a collection of axioms, theorems/conjectures, and related proofs, organized around some common theme, and the packages already cover an array of topics (not entirely independent of one another).
For convenience, each theory package also provides a collection of pre-defined common expressions and a ``demonstrations'' page to document the features of the package.
The current theory package hierarchy is illustrated in Figure \ref{fig:theory_package_hierarchy} (and can be accessed and interactively explored in depth at the \ProveItWebsite{}).

\begin{figure}[htb]
\captionsetup{width = 0.4\textwidth}
\centering
  \centering
  \footnotesize{
\begin{itemize}
\renewcommand\labelitemi{\scalebox{0.6}{ $\blacksquare$}}
\renewcommand\labelitemii{\small $\circ$}
\renewcommand\labelitemiii{\small $\circ$}
\renewcommand\labelitemiv{\small $\circ$}
    \item proveit.logic
    \begin{itemize}
        \item proveit.logic.booleans
        \begin{itemize}
            \item proveit.logic.booleans.implication
            \item proveit.logic.booleans.negation
            \item proveit.logic.booleans.conjunction
            \item proveit.logic.booleans.disjunction
            \item proveit.logic.booleans.quantification
            \begin{itemize}
                \item proveit.logic.booleans.quantification.universality
                \item proveit.logic.booleans.quantification.existence
            \end{itemize}
        \end{itemize}
        \item proveit.logic.equality
        \item proveit.logic.sets
        \begin{itemize}
            \item proveit.logic.sets.membership
            \item proveit.logic.sets.enumeration
            \item proveit.logic.sets.inclusion
            \item proveit.logic.sets.unification
            \item proveit.logic.sets.intersection
            \item proveit.logic.sets.subtraction
            \item proveit.logic.sets.comprehension
            \item proveit.logic.sets.disjointness
            \item proveit.logic.sets.cardinality
        \end{itemize}
    \end{itemize}
    \item proveit.numbers
    \begin{itemize}
        \item proveit.numbers.number\_sets
        \begin{itemize}
            \item proveit.numbers.number\_sets.naturals
            \item proveit.numbers.number\_sets.integers
            \item proveit.numbers.number\_sets.rationals
            \item proveit.numbers.number\_sets.real\_numbers
            \item proveit.numbers.number\_sets.complex\_numbers
        \end{itemize}
        \item proveit.numbers.numerals
        \begin{itemize}
            \item proveit.numbers.numerals.binaries
            \item proveit.numbers.numerals.decimals
            \item proveit.numbers.numerals.hexidecimals 
        \end{itemize}
        \item proveit.numbers.addition
        \begin{itemize}
            \item proveit.numbers.addition.subtraction 
        \end{itemize}
        \item proveit.numbers.negation
        \item proveit.numbers.ordering
        \item proveit.numbers.multiplication
        \item proveit.numbers.division
        \item proveit.numbers.modularity
        \item proveit.numbers.exponentiation
        \item proveit.numbers.summation
        \item proveit.numbers.product
        \item proveit.numbers.differentiation
        \item proveit.numbers.integration
    \end{itemize}
\end{itemize}
} 
\caption{A snapshot of the current \ProveIt{} theory package hierarchy, which can be accessed and interactively explored at www.pyproveit.org.  This conveys a tree-like structure.  At each level of the hierarchy, the order was chosen so that more primitive concepts come before more advanced concepts.  However, it is possible for a package to depend upon a later one (for example, some of the axioms/theorems in the \texttt{proveit.logic} package utilize number literals from the \texttt{proveit.number} package, but the number concept is generally more advanced than the logic concept).}
\label{fig:theory_package_hierarchy}
\end{figure}

\begin{figure*}[htb]
\captionsetup{width = 16cm}
\centering
  \includegraphics[width = 0.6\textwidth]{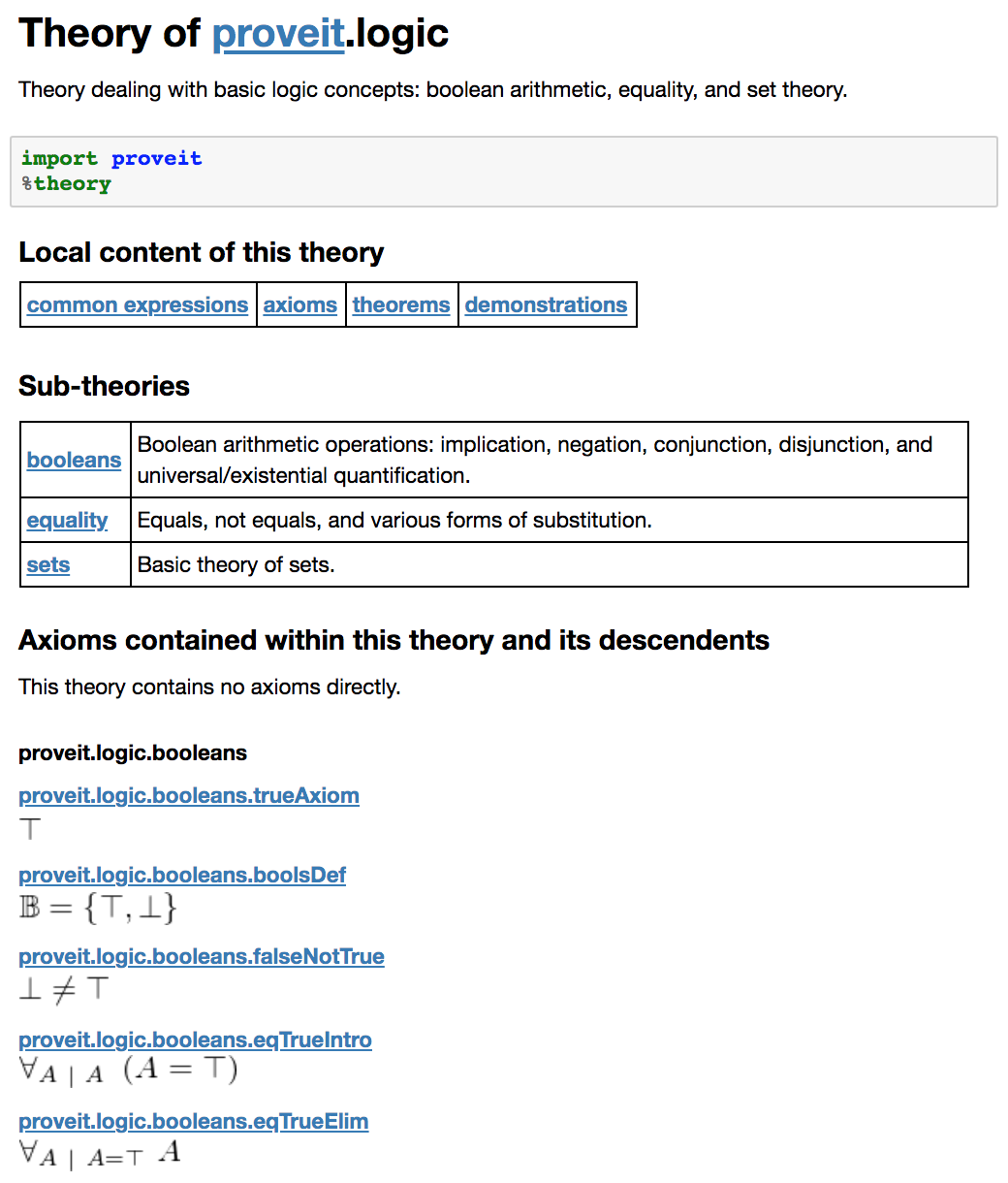}
\caption{A portion of the top-level theory notebook for the \texttt{proveit.logic} theory package, and an example of a ``theory of \ldots'' page in \ProveIt{}'s theory package system, with links to the package's notebooks for common\_expressions, axioms, theorems (and conjectures), and demonstrations, a listing of sub-theories, and the beginning of the list of all the axioms contained either at this theory level or at levels nested within. The full html version of this particular theory notebook can be viewed at \url{http://pyproveit.org/packages/proveit/logic/_theory_nbs_/theory.html}.}
\label{fig:example_theory_nb}
\end{figure*}

Appendix~\ref{sec:Axioms} lists some of the fundamental theory packages of the current system and their axioms. For example, one very broad theory package in \ProveIt{} is the \texttt{logic} package, inside of which appears the \texttt{sets} package, which itself contains a growing number of sub-packages devoted to topics such as \texttt{cardinality}, \texttt{membership}, \texttt{enumeration}, \textit{etc}. Part of the top-level notebook for the \texttt{proveit.logic} theory package appears in Figure~\ref{fig:example_theory_nb}, showing the links to its common expressions, axioms, theorems, and demonstrations notebooks, as well as links to its sub-theories and the first few items in its list of contained axioms.\footnote{We use \textit{theory packages} for organizational purposes to encapsulate axioms and related theorems. Such theorems, however, can (and often do) utilize, in their proofs, axioms and theorems from external packages. Thus a theory package does not generally constitute a ``formal theory'' in the strict sense of containing only theorems derivable from the specific axioms of the package itself.}

\begin{figure*}[htb]
\captionsetup{width = 16cm}
\centering
  \includegraphics[width = 0.7\textwidth]{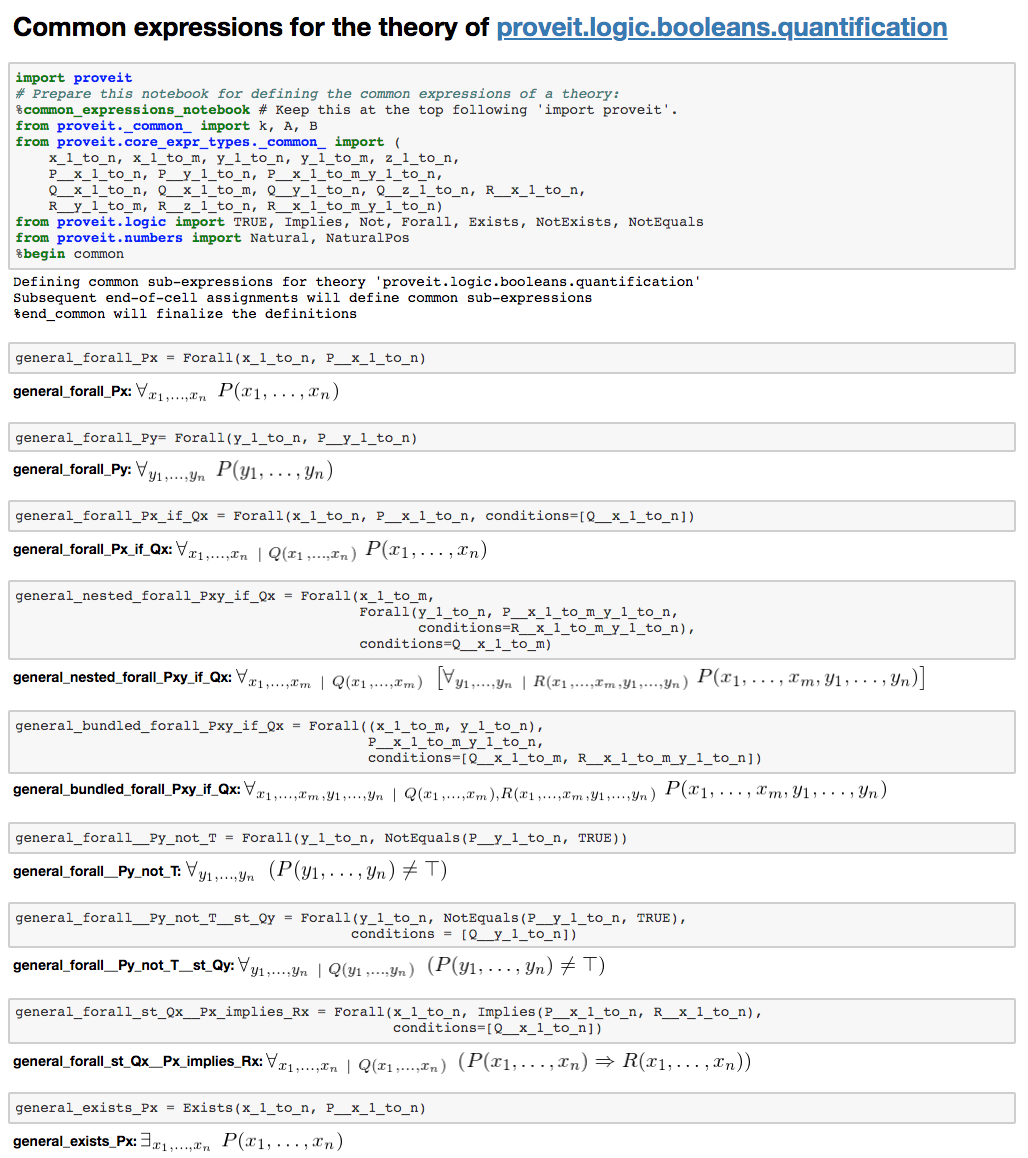}
\caption{A screenshot of the ``common expressions'' notebook in \ProveIt{}'s \texttt{logic.booleans.quantification} theory package, and an example of the many such ``common expressions'' notebooks available throughout the \ProveIt{} theory package system. Such notebooks are easily amended and provide useful predefined expressions that can be easily imported into proof notebooks, saving time and effort.  The full html version of this particular common expressions notebook can be viewed at \url{http://pyproveit.org/packages/proveit/logic/booleans/quantification/_theory_nbs_/common.html}.}
\label{fig:example_common_nb}
\end{figure*}

\begin{figure}[htb]
\captionsetup{width = 8cm}
\centering
  \includegraphics[width = 0.45\textwidth]{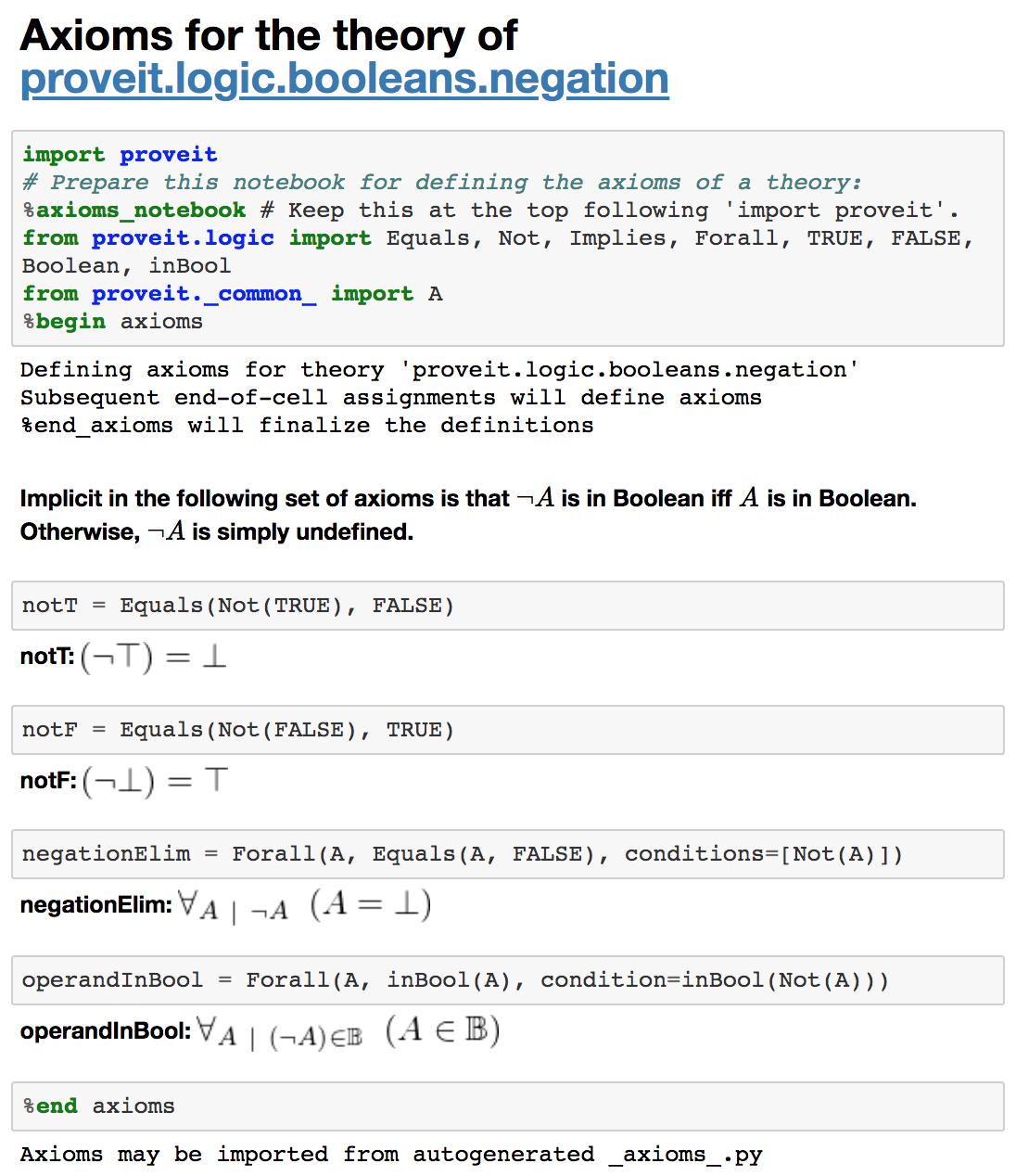}
\caption{The axioms notebook in the \texttt{logic.booleans.negation} theory package. Axiom notebooks appear throughout the \ProveIt{} theory package hierarchy, typically providing axiomatic definitions on which the package is based. See Appendix \ref{sec:Axioms} for an extensive listing of package axioms.  The html version of this particular axioms notebook can be viewed at \url{http://pyproveit.org/packages/proveit/logic/booleans/negation/_theory_nbs_/axioms.html}.}
\label{fig:example_axioms_nb}
\end{figure}

\begin{figure*}[htb]
\captionsetup{width = 16cm}
\centering
  \includegraphics[width = 0.6\textwidth]{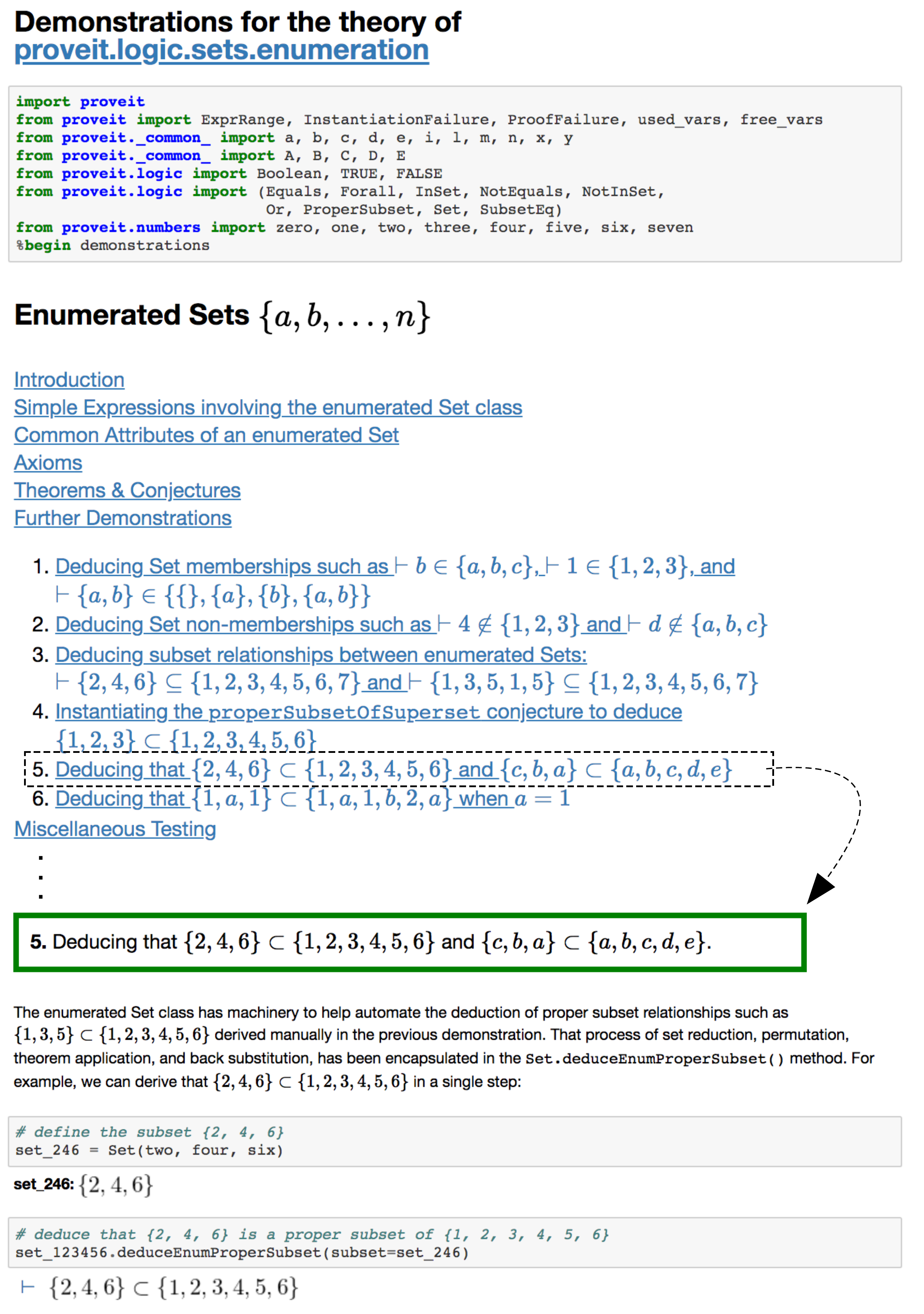}
\caption{Two portions of the demonstrations notebook found in the \texttt{logic.sets.enumeration} theory package, showing the table of contents (the vertical ellipsis indicates omitted content) and the beginning of one of several demonstrations included in the notebook, in this case demonstration \#5 illustrating how to define an enumerated set and prove that one set is a proper subset of another under appropriate assumptions. The full html version of this particular demonstrations notebook can be viewed at \url{http://pyproveit.org/packages/proveit/logic/sets/enumeration/_theory_nbs_/demonstrations.html}.}
\label{fig:example_demonstrations_nb}
\end{figure*}

We demonstrate a common expressions notebook in Figure \ref{fig:example_common_nb},
for the {\small\texttt{logic.booleans.quantification}} theory package, 
an axioms notebook in Figure \ref{fig:example_axioms_nb}, for the\\ {\small\texttt{logic.booleans.negation}} theory package, 
and the demonstrations notebook in Figure~\ref{fig:example_demonstrations_nb}, from the\\ {\small\texttt{logic.sets.enumeration}} theory package.  The common expressions, axioms, theorems, and demonstrations pages are all Jupyter notebooks that display judgments, expressions, and proofs using \LaTeX-formatting. Each one is also available as an html web page on the \ProveItWebsite{}.  Additionally, a theory package has python scripts which define expression classes and useful code for invoking theorems and performing various degrees of proof automation. 

\subsection{Axioms}

An \textbf{\textit{axiom}} in \ProveIt{}, consistent with the commonly accepted meaning of the term, consists of a judgment of the form $\vdash C$, with no assumptions and no free variables, and which, by its definition and role, has no associated proof. As is common in mathematics and logic generally, axioms in \ProveIt{} often provide the defining properties of various expression types, and can easily be ``posited'' by the \ProveIt{} user through an interactive construction process (either by adding axioms to an already-existing theory package or by creating an entirely new package within the \ProveIt{} TPS).

For example, the following transitivity, reflexivity, and symmetry axioms appear in the \texttt{logic/equality} context:
\begin{align}
    &\vdash \forall_{x, y, z \rvert x=y, y=z} [x = z],\\
    &\vdash \forall_{x} [x = x],\\
    &\vdash \forall_{x, y} [(y=x) = (x=y)],
\end{align}
\noindent and in the \texttt{logic/boolean/implication} context we have the axiomatic definition of logical equivalence:
\begin{align}
    &\vdash \forall_{A, B} [(A\Leftrightarrow B) = \left((A\Rightarrow B) \land (B\Rightarrow A)\right)].
\end{align}
As another example, a screenshot of the entire axiom notebook for the \texttt{logic.booleans.negation} package is shown in Figure \ref{fig:example_axioms_nb}.

Axioms will often appear as leaves in proof DAGs and can be explicitly imported into other package notebooks to support user explorations and interactive theorem-proving. 

\subsection{Theorems, Conjectures, and Dependencies}

\noindent \textbf{\textit{Theorems}} in \ProveIt{} are similar in many ways to axioms.  They also are judgments with no assumptions and no free variables, and they are grouped into theory packages.  Unlike axioms, however, theorems need proofs.  A theorem without a proof in the system is marked as a \textbf{\textit{conjecture}}.  It may still be used in the proof of other theorems, but those theorems will also be marked as conjectures (in \ProveIt{} theorem notebooks, such a conjecture is marked as a ``conjecture with conjecture-based proof''). A conjecture converts to a theorem when it has a proof in the \ProveIt{} system and the proof depends only upon axioms and other fully proven theorems (in a non-circular manner, of course). This easy availability of conjectures provides valuable flexibility, allowing top-down proof exploration and development.  Users can generate interesting proofs based upon ``conjectures'' whose proofs can wait.  Often, these are not conjectures in the true sense; rather, they are well-established facts that derive, in principle, from proper axioms, but simply have not yet been proven in the \ProveIt{} system.  We use this feature a lot.  It takes a substantial amount of work to build up a breadth of knowledge from only the most fundamental facts (axioms) and it can often be more valuable to prioritize theorem-proving of sophisticated theorems from which there is more insight to be gained.  Of course, this feature could also be used for ``authentic'' conjectures (e.g., you could prove an algorithm complexity theorem based upon the conjecture that $\textrm{P} \neq \textrm{NP}$).

Each theorem has an associated ``proof'' notebook (a Jupyter notebook in the interactive system or a web page in the \ProveItWebsite{}).  The proof of the theorem may or may not be populated.  When the proof is completed, the last cell will be a ``\texttt{\%qed}'' command with the proof DAG as the output.  An example of such a proof notebook for $\sqrt{2} \notin \mathbb{Q}$ is shown in \S\ref{Sec:Example}.  It may, however, depend upon other theorems/conjectures.\footnote{It is worth noting that \ProveIt{} makes no distinction between a theorem and a lemma. What a user might consider to be a lemma, \ProveIt{} considers just another theorem with no special status, although the Jupyter notebook user-interface will allow the user to include contextual comments and explanations if desired.}

While theorems/conjectures may be imported and instantiated directly in the construction of a proof (for a different theorem or a stand-alone proof), the best practice is to write methods in the Python scripts of the theory package that will instantiate these theorems indirectly.  That way, the details of the theorem instantiation (the name of the theorem and choosing the expressions for instantiating each variable) are encapsulated in the package. Furthermore, \ProveIt{} has useful automation features in which new facts are derived from other facts as they are created (e.g., when $S \vdash x = y$ is proven, $S \vdash y = x$ is automatically derived as a side-effect) or are automatically concluded as needed (e.g., given $S \vdash A$ and $T \vdash B$, $S \cup T \vdash A \land B$ can be concluded automatically). 
Ideally, constructing the proof then becomes, after some experience and gaining familiarity with the theory packages involved, as natural as writing an informal proof by hand but with certainty of correctness and less tedious writing.  Our example in \S\ref{Sec:Example} has many instances of theorems being instantiated indirectly, either through full automation (see the cells that simply call the \texttt{prove} method) or convenient methods that perform directed transformations.

\subsubsection{Theorem dependencies form a DAG}
\label{Sec:TheoremDependencies}

Just as there must not be any circular dependency within a particular proof object, there must not be any circular dependency among the various theorem proofs.  That is, a theorem may have a proof that depends upon other theorems  whose proofs depend upon other theorems, and so forth.  Individual proofs may also depend upon axioms and unproven conjectures.  These dependencies form a graph that must be acyclic to avoid circular logic at the theorem level.  The leaf nodes are axioms or conjectures.  With a particular theorem at the root, when there are only axioms as leaf nodes and no conjectures, that theorem is said to be ``fully proven'' and is no longer marked as conjecture.

Because theorems may be proven in any order and may be invoked indirectly via automation, \ProveIt{} requires a mechanism to constrain which theorems are ``usable'' in any given theorem proof.  For example, say that it is intended for theorem $B$ to depend upon theorem $A$ but neither have been proven.  However, automation has already been developed for automatically instantiating theorem $B$ as needed.  Furthermore, assume that this automation would  allow theorem $A$ to be trivially but erroneously proven as a consequence of theorem $B$.  That would create a problem because if we let theorem $A$ be proven using theorem $B$ as a dependency, then theorem $B$ would not be allowed to use theorem $A$ as intended (since the dependencies must be acyclic).  Our solution is to explicitly indicate the theorems and packages that may be \textit{presumed} within each theorem proof.  In cell 2 of our $\sqrt{2} \notin \mathbb{Q}$ demonstration of \S\ref{Sec:Example}, shows an example of ``presuming'' theory packages.\footnote{We intend to make our ``presuming'' machinery more convenient for users/developers, so these details are subject to change.}

For a conjecture with a proof, it is useful to know what dependent conjectures remain to be proven for it to gain the ``fully proven'' status.  For any theorem/conjecture with a proof, it is useful to know what axioms are required (as leaves of the dependency DAG).  After all, it is easy to add axioms, so it is a good idea to make sure that the axioms used in a proof of interest are trusted.\footnote{To really be certain about a proof, one should examine the expression DAG structure of each axiom as well as the proven theorem.}  Going the other direction, it can also be useful or interesting to know all of the theorems that depend upon an axiom or theorem.  All of this information is presented on ``dependency'' notebooks (a Jupyter notebook in the interactive system or a web page in the \ProveItWebsite{}).  There are hyperlinks to \texttt{<dependencies>} on each ``proof'' notebook as well as the ``expression information'' notebooks that may be reached by clicking on any axiom / theorem / conjecture expression.  Stripped down contents of the ``dependency'' notebook for the $\sqrt{2} \notin \mathbb{Q}$ example is shown in \S\ref{Sec:ExampleDependencies} (it excludes the names of the axioms / conjectures that appear in the actual ``dependency'' notebook and is displayed in a more compact form).


\section{Example problem: \texorpdfstring{$\sqrt{2} \notin \mathbb{Q}$}{sqrt(2) not in Rationals}}
\label{Sec:Example}

In this section we present a ``theorem proof'' notebook for $\vdash \sqrt{2} \notin \mathbb{Q}$.  This is a nice example that is not too trivial but not terribly complicated. Furthermore, we can compare our demonstration with a convenient compilation of this same demonstration implemented in 17 other theorem provers~\cite{Wiedijk:2006_17_Provers}.  Two additional proofs stand out among the others as easily comprehensible by anybody who is sufficiently well-versed in mathematics: a geometrical proof and an informal proof.  However, these are not implemented in a formal language that can be verified through automation.  In the formal proof demonstrations, the strategies employed to generate the proof are not obvious without being an expert in that particular system.

Our proof notebook is shown in the following pages.  As with all Jupyter notebooks, there are input cells (labeled ``\texttt{In[\#]}'') and corresponding output cells (labeled ``\texttt{Out[\#]}'').  The inputs show our commands for constructing the proof and the outputs show the progress along the way (usually a proven judgment whose proof is accessible via the hyperlinked $\vdash$ symbol).  Additionally, there are ``markdown'' cells that describe the intentions as we work through the proof.  The final cell issues a \texttt{\%qed} command followed by the output of the generated proof DAG.  Our proof has generated, largely through computer automation, a proof DAG with 155 steps (nodes), but we only show a few of these in this paper for brevity.\footnote{The full version can be viewed at
\url{http://pyproveit.org/packages/proveit/numbers/exponentiation/_theory_nbs_/proofs/sqrt2_is_not_rational/thm_proof.html}.}
Some of these steps invoke theorems/conjectures.  Currently, this proof still has a ``conjecture'' status since we have not proven all of the dependent conjectures.  The list of dependent conjectures/axioms (the leaf nodes of the dependency DAG) is shown in \S\ref{Sec:ExampleDependencies}.

We hope that our proof stands out among the $\sqrt{2} \notin \mathbb{Q}$ proofs of other formal systems~\cite{Wiedijk:2006_17_Provers} as being nearly as straightforward and readable as the informal proof.  The output judgments are easily readable by anybody familiar with the mathematical notation.  The proof DAG is similarly readable with some basic understanding of our inference rules~[Appendix~\ref{sec:InferenceRules}].  Also, it is not hard to deduce the purpose and effect of the input commands, given the ``markdown'' comments and the output.  Therefore, one does not need to be a \ProveIt{} expert in order to understand a proof construction and proof DAG in a \ProveIt{} notebook.  Constructing a proof does require a certain amount of expertise and practice, but it certainly helps to have comprehensible demonstrations as examples;  many others can be found on the \ProveItWebsite{}.  Automation features and convenient methods for applying other theorems/conjectures of the system are also tremendously valuable.  This example demonstrates a number of these features/methods.  Our ultimate goal is to make it as easy (or easier) to construct a formal proof as it is to construct an informal but rigorous one.  With sufficient theory package development, the complexity of the mathematics should not be a barrier to this goal.

\clearpage
\includepdf[pages=-,pagecommand={},width=\textwidth]{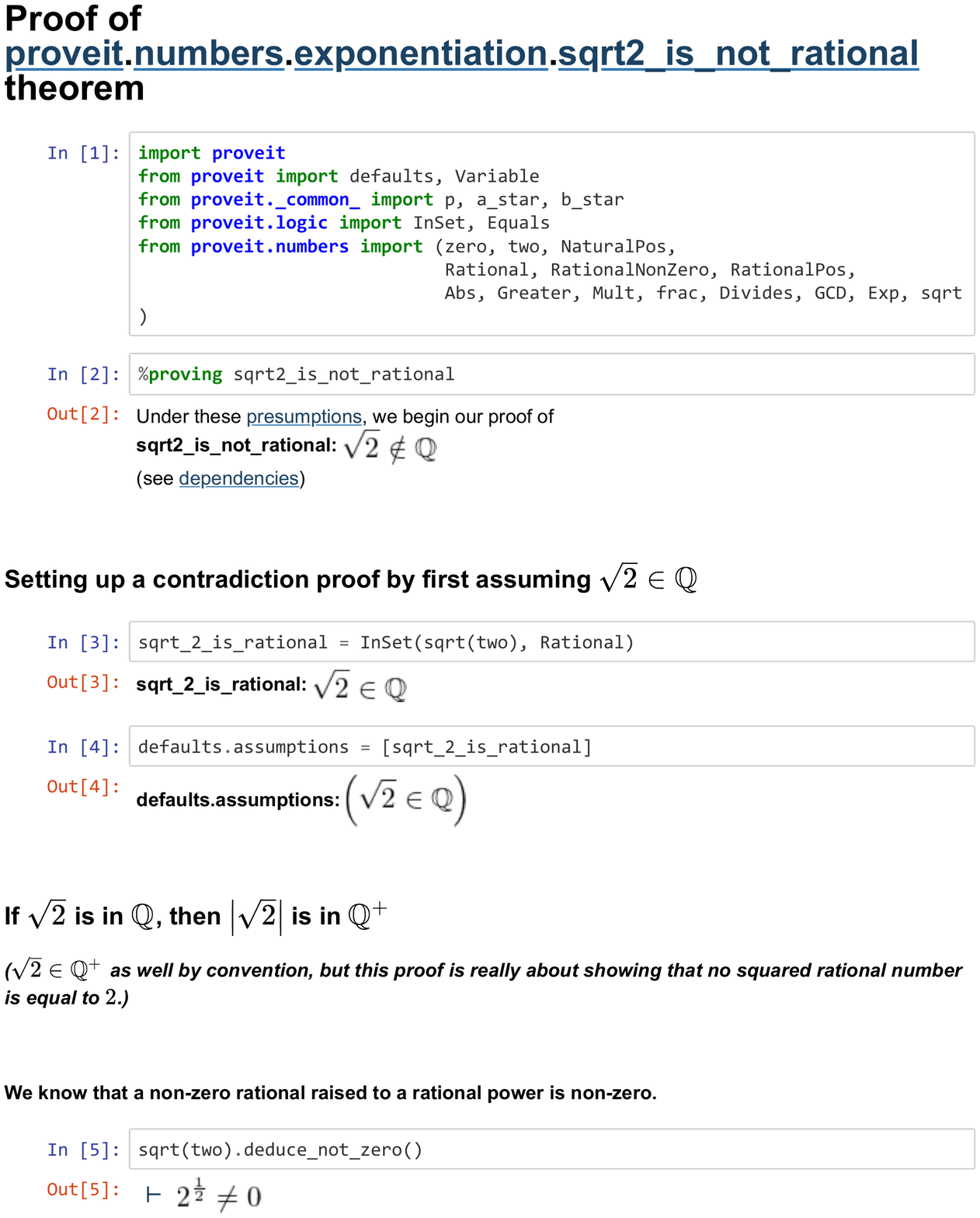}
\clearpage

\subsection{Dependencies for our \texorpdfstring{$\sqrt{2} \notin \mathbb{Q}$ proof}{sqrt(2) not in Rationals}}

\input{example_dependencies}
\label{Sec:ExampleDependencies}


\section{Avoidance of known paradoxes}
\label{Sec:Paradoxes}

At this time, we do not have a full consistency proof for the logic and basic axioms of \ProveIt{}.  Even before this is achieved, \ProveIt{} can be a valuable tool.  Its proofs are designed to be human-readable and well-structured, and therefore should be at least as acceptable as very rigorous proofs generated by hand that are the standard in much of the literature.  That is, you can take its proofs at face value for experts to decide if the arguments, specific to the proof, are sound.  Still, one would hope that it is not possible to derive
\begin{align*}
    \vdash\bot
\end{align*}
using standard axioms of \ProveIt{}.  A consistency proof would serve the purpose of arguing that this is not possible.  While we are not prepared to claim that we can rigorously prove this to be the case just yet, we are prepared to discuss some specific well-known logical fallacies and how \ProveIt{} does not fall into their traps.

\subsection{Russell's paradox (what is a set?)}
\label{Sec:RussellsParadox}
\input{Russells_paradox}

\subsection{Kleene-Rosser-Curry’s paradox (what is a function?)}
\label{Sec:CurrysParadox}
\input{currys_paradox}


\section{Discussion \& Conclusions}
\label{Sec:Conclusion}

We introduce \ProveIt{}, a Python-based general-purpose interactive theorem-proving assistant designed with the goal to make formal theorem proving as easy and natural as informal theorem proving.  The Jupyter notebook-based interface and underlying Python code make the \ProveIt{} system accessible to a wide audience, moderately easy to learn and use, and relatively easy to make contributions to its expandable knowledge base of axioms, conjectures, theorems, proofs, and proof tactics.

\ProveIt{} expressions, proofs (that may depend upon theorems and conjectures), and proof dependencies are all represented internally by directed acyclic graphs (DAGs).  Proof and proof dependency DAGs help organize a proof result into a human-readable presentation and ensure against circular logic.  Expression DAGs make the connection between internal  representations used by the core logical system and mathematical notation formatted elegantly in \LaTeX{}.  By supporting the use of conjectures, sophisticated and interesting theorems may be tentatively proven, dependant upon obvious or commonly accepted facts before they are fully proven in the system.  Details of our system are provided in the appendices.

As an example, we show a proof of $\sqrt{2}\notin\mathbb{Q}$ which demonstrates many of \ProveIt{}'s features.  We show how, with the use of automation and methods (proof tactics) that encapsulate axiom/theorem/conjecture invocation, the construction of a formal proof in \ProveIt{} does, in fact, resemble an informal proof.  We demonstrate the use of conjectures that allow us to present such a demonstration before we have had the chance to prove some basic, simple facts that are needed for this proof.

We present arguments in support of the logical consistency of \ProveIt{}, addressing two well known potential pitfalls in particular: Russell's paradox and Curry's paradox.  We believe that we avoid these pitfalls but do acknowledge that we lack a rigorous consistency proof at this time.  Our system uses nonstandard approaches in its avoidance of requiring type theory that merits further investigation by experts in mathematical foundations.  However, because our proofs are presented in an intuitive, human-readable form, the value of the system is not substantially diminished by the lack of complete certainty regarding its consistency.  Our system makes it relatively easy to generate formal, structured proofs, with no gaps in the chain of reasoning, that may be examined at face value by experts who can decide whether or not a given proof is convincing.

\begin{figure*}[hbt]
\begin{center}
\scalebox{0.55}{\includegraphics{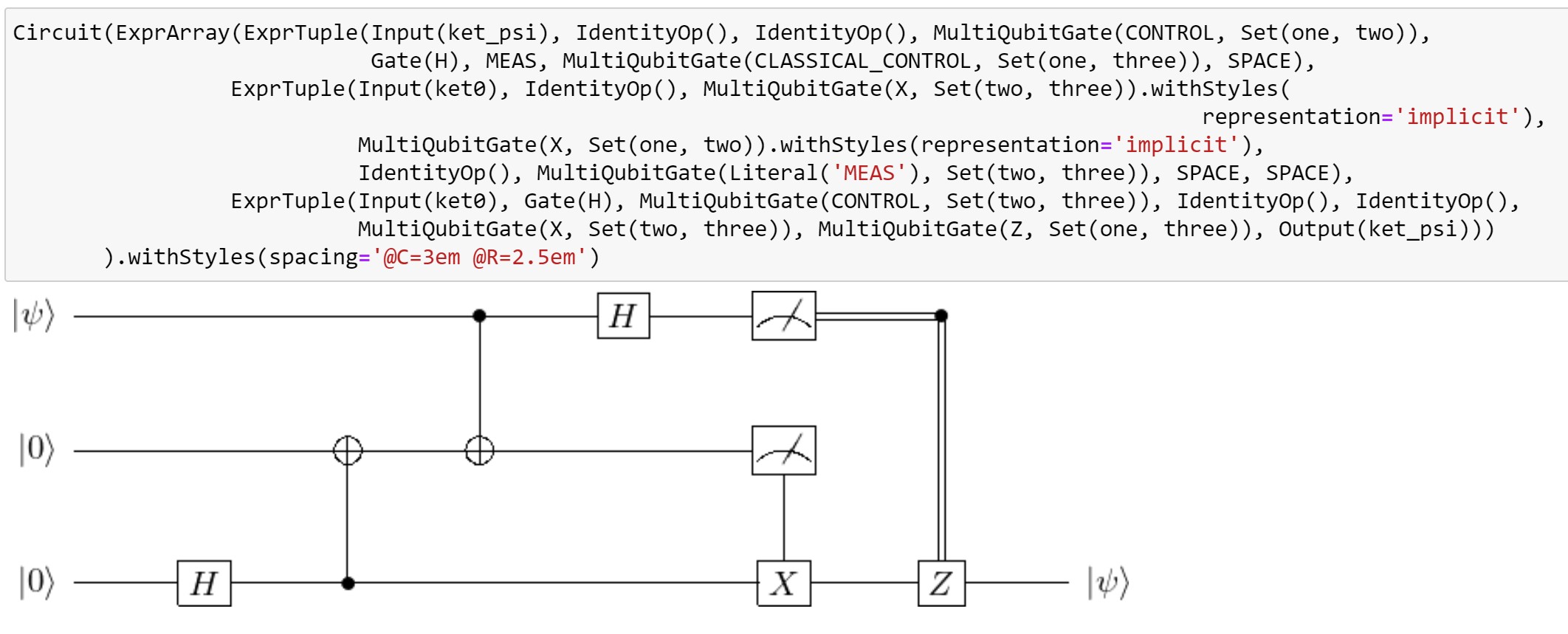}} \caption{An example of a quantum circuit generated using Prove-It (in this case, representing a quantum teleportation algorithm). }\label{fig:example_quantum_circuit}
\end{center}
\end{figure*}

\begin{figure*}[hbt]
\begin{center}
\scalebox{0.7}{\includegraphics{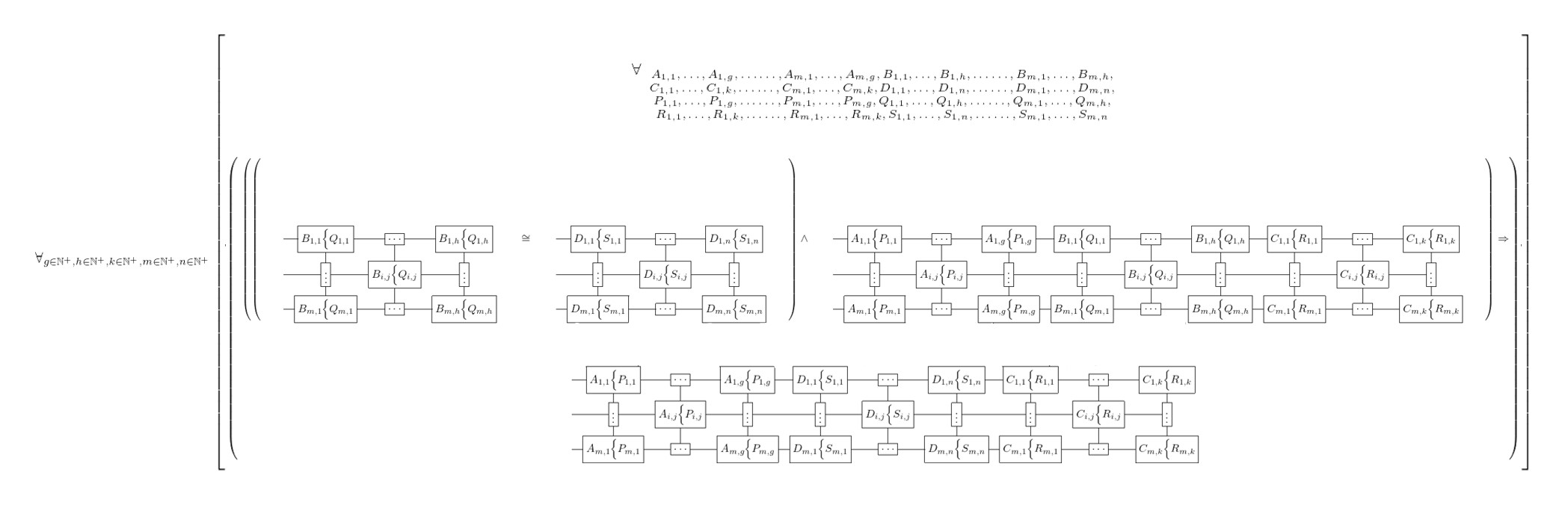}} \caption{Our temporal\_circuit\_substitution theorem that allows us to replace an equivalent portion of a valid (true) circuit of arbitrary size. The theorem states that for any valid quantum circuit, labeled in a manner that expresses a true identity analogous to an equation, the replacement of a portion of this circuit with an equivalent circuit of the same size results in another valid circuit.  This is done with the use of generic multi-qubit gates and Prove-It's ExprRanges.} \label{fig:time_equiv_judgment_thm}
\end{center}\end{figure*}

In addition to ongoing development and expansion of fundamental theory packages in logic, set theory, numerical sets, \textit{etc}, current development and future work involves the application of \ProveIt{} to issues of quantum computation and quantum circuit verification and design. As a quick preview of such work, Figure \ref{fig:example_quantum_circuit} shows an example quantum circuit (representing a quantum teleportation algorithm) generated in the \ProveIt{} system, and Figure \ref{fig:time_equiv_judgment_thm} shows a quantum-circuit-based expression of a general theorem in \ProveIt{} for replacing an equivalent portion of a valid quantum circuit of arbitrary size.  Future development may also include interfacing with a SAT or SMT solver and generally strengthening its automation capabilities.

\section{Acknowledgments}
We gratefully acknowledge Deepak Kapur, Kenny Rudinger, Mohan Sarovar, Jon Aytac, Geoffrey Hulette, Denis Bueno, and Geoffrey Reedy for valuable discussions and contributions to the development of \ProveIt{}.

This work was supported by the the U.S. Department of Energy, Office of Science, Office of Advanced Scientific Computing Research under the Quantum Computing Applications Team (QCAT) program, and the Laboratory Directed Research and Development program at Sandia National Laboratories. Sandia National Laboratories is a multi-program laboratory managed and operated by National Technology and Engineering Solutions of Sandia, LLC., a wholly owned subsidiary of Honeywell International, Inc., for the U.S. Department of Energy's National Nuclear Security Administration under contract DE-NA-0003525.  This paper describes  objective  technical  results  and  analysis.   Any subjective views or opinions that might be expressed in the paper do not necessarily represent the views of the U.S. Department of Energy or the United States Government.

\vspace{0.5in}

\clearpage

\appendix
\appendixpage
\begin{appendices}
\section{Core/primitive Expression Classes}\label{appendix_ExprTypes}
\label{Sec:ExprTypes}
\input{core_types_and_literals}

\section{Expression style choice}\label{appendix_Style}
\label{Sec:Style}
\input{styles}

\section{Inference Rules}\label{appendix_InferenceRules}
\label{sec:InferenceRules}
\input{inference_rules}

\section{Reduction Rules}\label{appendix_ReductionRules}
\label{sec:ReductionRules}
\input{reduction_rules}

\section{Basic Theories and Axioms}
\label{sec:Axioms}
\input{axioms.tex}

\end{appendices}
\clearpage

\vspace{0.5in}


\bibliographystyle{unsrt}
\bibliography{bibliography.bib}

\end{document}

%% file: DomainsOfDiscourse.tex
In deductive systems, it is typical to either presume a domain of discourse or use a type system which determines the type of any expression (term).  In propositional logic~\cite{Mendelson:2015}, the domain of discourse is presumed to be the set of Boolean values (true or false which we denote as $\top$ and $\bot$ respectively).  That is, every variable may be interpreted as being either true or false, and propositional formulas have a certain structure (well-formedness) to ensure they may only be true or false.  In Zermelo–Fraenkel set theory\cite{Jech:2003_Set_Theory}, the domain of discourse is presumed to be sets.  That is, every variable may be interpreted as being a set which is identified by its membership properties (\textit{i.e.},  what elements does the set contain?).  In a typed system, if $x$, $y$, and $z$ are integers, $x \cdot y - z$ can be determined to be an integer independent from the derivation logic of the system.  In contrast, \ProveIt{} never presumes a domain of discourse and does not use a type system.  Instead, domain restrictions in \ProveIt{} must be expressed as explicit assumptions or conditions, or must be derived from explicit assumptions or conditions in order to be utilized.

To be more precise and concrete, let us first introduce \emph{judgments} as the basic building blocks of any proof in our system.  Borrowing from the language of natural deduction proof calculi~\cite{Pelletier:1999_History_of_Natural_Deduction, Prawitz:1965_Natural_Deduction, Prawitz:2006_Natural_Deduction, Martin-Lof:1996}, a \emph{judgment} is an object of 
knowledge, presented in the form
\begin{equation}
    \{A_1, \ldots, A_n\} \vdash B
\end{equation}
for some natural number $n$.
The meaning of such a judgment is that $B$ is true if $A_1$ through $A_n$ are each true.  
``$\vdash$'' is known as the turnstile symbol and represents logical entailment or deducibility.  Essentially, if this judgment is true, it means that $B$ follows logically if we assume that $A_1$ through $A_n$ are each true.
We call $A_1$ through $A_n$ the assumptions of this judgment.  We use curly braces to signify that the order and redundancy
of the assumptions is unimportant (essentially acting as an enumerated, unordered set which is 
typically denoted with curly braces).  When there are no assumptions (\textit{i.e.}, $n=0$), we simply write this with nothing 
on the left side of the turnstile.  For example,
\begin{equation}
    \vdash B
\end{equation}
means that $B$ can be derived to be true without any assumptions.

In the next section, we will discuss \emph{expressions} as the building blocks of judgments and we will describe how a proof is constructed using the judgment building blocks.  In this section, we focus on the meaning of the judgments themselves at a meta-logical level, in relation to domains of discourse in particular.  In \ProveIt{}, the assumptions ($A_i$) and consequent ($B$) do not need to be Boolean-valued.  For the sake of the judgment, we only care about whether or not the parts of the judgment are true.  We regard every expression, even if it contains gibberish, to be either equal to true or not equal to true.\footnote{Note, not being equal to true is not the same as being false.  For example, the number 1 is not equal to true, but is not false either.  Also, it may not always be possible to know whether or not any given expression is equal to true; that is, \ProveIt{} is not a ``complete'' logic system and there is no intention to make it complete.}  The translation of $\{A_1, A_2\} \vdash B$ is quite literally, ``we can derive $B$ to be equal to true if we assume that $A_1$ and $A_2$ are both equal to true.''  This says nothing about $A_1$, $A_2$, and $B$ having to be genuine propositions (unlike natural deduction systems, for example, in which you must first prove the parts of the judgment to be well-formed propositions before you can prove them to be true).  If an assumption is used that is not a genuine proposition, this will simply be a matter of ``garbage-in, garbage-out.''  What we care about is truth, and anything that is or can be legitimately true will necessarily be a perfectly legitimate proposition.
Ultimately, this is only a matter of language, but our unique approach frees us from needing a type system and is always explicit about the domains of variables (via assumptions, conditions, or derivations).

Let us present some examples.  Consider the ``law of excluded middle.''  No problem.  We simply need to be explicit about variables being in the Boolean domain (which we denote with $\mathbb{B}$).\footnote{See Table~\ref{tab:symbols} for a list of logic and set theory symbols.}
\begin{equation}
    \{A \in \mathbb{B}\} \vdash A \lor \lnot A
\end{equation}
and
\begin{equation}
    \{A \in \mathbb{B}, B \in \mathbb{B}\} \vdash (A \land B) \lor \lnot (A \land B)
\end{equation}
can each be proven, but the explicit assumptions are important.  We can also prove harmless things about gibberish.  For example,
\begin{equation}
    \{x = (5 \lor \top), x + 10\} \vdash (5 \lor \top) + 10.
\end{equation}
In this example, we are mixing types ($5$ and $10$ are numbers and $\top$ is a Boolean) in a way that is meaningless to mathematicians and we are concluding something that is clearly not a proposition.  It does not matter for our purpose of proving judgments.  If we assume that the assumptions on the left are true statements, we can derive the consequent on the right.  Garbage in, garbage out.  If you find such garbage to be offensive, it would be easy, in principle, to filter it out.  By allowing ``garbage,'' however, we avoid the need of a type system that can feel clunky to those who are not experts in formal methods.

Our lack of implicit Boolean type restrictions extends to implications and conditions of quantifiers.  In \ProveIt{}, the implication $A \Rightarrow B$ means ``$B$ equals true if $A$ equals true.''\footnote{$A \Rightarrow B$ has a similar meaning to the judgment $\{A\} \vdash B$.  $\{A\} \vdash B$ may be derived from $\vdash A \Rightarrow B$ and vice-versa.  However, $\{A\} \vdash B$ means that $B$ is deducible when assuming $A$ which depends upon the completeness of the theory while $A \Rightarrow B$ has a truth value that depends only upon truth values of $A$ and $B$.  
Furthermore, we prove judgments, not implications (directly).  An implication may be used within a judgment and may be nested within another implication, but a judgment may not be nested within another judgment.}  Therefore,
\begin{equation}
    \vdash 3 \Rightarrow 3,
\end{equation}
is a perfectly valid judgment that simply means $3$ is equal to true if we assume it to be so.  We could also prove $\vdash 3 \Rightarrow 5$ if we included axioms to prove that Booleans are distinct from numbers.\footnote{We are currently agnostic about whether or not Booleans and numbers are disjoint sets.  We embrace agnosticism regarding things that are irrelevant.}  Furthermore, $\forall_{x~|~Q(x)}~P(x)$ means ``$P(x)$ is true for any $x$ for which $Q(x)$ is true.''  Therefore,
\begin{equation}
    \vdash \forall_{A~|~\lnot A}~\lnot(A \land \top)
\end{equation}
is valid even without an explicit restriction on the domain of $A$.
Our quantifier has a condition $\lnot A$ from which $A \in \mathbb{B}$ can be derived 
according to the current axioms of our system [Appendix \ref{sec:Axioms}].
Specifically, we can only prove that $\lnot A$ is true if $A$ is false.  It does not matter that $\lnot 5$ is gibberish, only that $\lnot 5 \neq \top$.  Since ``false'' is the only value of $A$ for which $\lnot A$ is true, and since $\lnot(A \land \top)$ is true if $A=\bot$, this is a provable and valid judgment by our definitions.

%% file: example_dependencies.tex
Unproven conjectures required (directly or indirectly) to prove \texttt{sqrt2\_is\_not\_rational}
\begin{itemize}
\item $\forall_{i \in \mathbb{N}}~\left[\forall_{f}~\left(|\left(f(1), \ldots, f(i)\right)| = |\left(1, \ldots, i\right)|\right)\right]$
\item $\forall_{a \in \mathbb{Q}^{\neq 0}}~\left(\left|a\right| \in \mathbb{Q}^+\right)$
\item $\forall_{a, b \in \mathbb{N}^+~|~gcd(a, b) = 1}~\left[\forall_{p \in \mathbb{N}^+~|~p > 1}~(\lnot \left(\left(p \rvert a\right) \land \left(p \rvert b\right)\right))\right]$
\item $\forall_{k \in \mathbb{Z}, a \in \mathbb{Z}, n \in \mathbb{N}^+~|~k \rvert a}~\left(k^{n} \rvert a^{n}\right)$
\item $\forall_{a, b, k \in \mathbb{C}~|~\left(k \cdot a\right) \rvert \left(k \cdot b\right), k \neq 0}~\left(a \rvert b\right)$
\item $\forall_{a, n \in \mathbb{Z}~|~2 \rvert a^{n}}~\left(2 \rvert a\right)$
\item $\forall_{x \in \mathbb{C}, y \in \mathbb{Z}~|~x \neq 0}~\left(x \rvert \left(x \cdot y\right)\right)$
\item $\forall_{a, b \in \mathbb{Q}^{\neq 0}}~\left(\frac{a}{b} \in \mathbb{Q}^{\neq 0}\right)$
\item $\forall_{x \in \mathbb{C}}~\left(\frac{x}{1} = x\right)$
\item $\forall_{a, b, c, d, e \in \mathbb{C}~|~c \neq 0}~\left(\left(\frac{a}{b \cdot c} \cdot \frac{c \cdot d}{e}\right) = \left(\frac{a}{b} \cdot \frac{d}{e}\right)\right)$
\item $\forall_{a \in \mathbb{N}^+, b \in \mathbb{N}}~\left(a^{b} \in \mathbb{N}^+\right)$
\item $\forall_{a, b \in \mathbb{Q}^{\neq 0}}~\left(a^{b} \neq 0\right)$
\item $\forall_{n \in \mathbb{N}^+, x \in \mathbb{R}^+}~\left((x^{\frac{1}{n}})^{n} = x\right)$
\item $\forall_{a \in \mathbb{C}, b \in \mathbb{C}, n \in \mathbb{N}^+}~\left(\left(a \cdot b\right)^{n} = \left(a^{n} \cdot b^{n}\right)\right)$
\item $\forall_{a \in \mathbb{Q}}~\left(\left|a\right|^{2} = a^{2}\right)$
\item $\forall_{x \in \mathbb{C}}~\left(x^{2} = \left(x \cdot x\right)\right)$
\item $\forall_{x \in \mathbb{C}}~\left(\left(1 \cdot x\right) = x\right)$
\item $\forall_{x \in \mathbb{C}}~\left(\left(x \cdot 1\right) = x\right)$
\item $\forall_{a, x, y \in \mathbb{C}~|~x = y}~\left(\left(x \cdot a\right) = \left(y \cdot a\right)\right)$
\item $\mathbb{R} \subset \mathbb{C}$
\item $\mathbb{N} \subset \mathbb{Z}$
\item $\mathbb{N}^+ \subset \mathbb{N}$
\item $\mathbb{Z} \subset \mathbb{Q}$
\item $\mathbb{N}^+ \subset \mathbb{Q}^{\neq 0}$
\item $\forall_{q \in \mathbb{Q}~|~q \neq 0}~\left(q \in \mathbb{Q}^{\neq 0}\right)$
\item $\mathbb{Q}^+ \subset \mathbb{Q}$
\item $\forall_{q \in \mathbb{Q}^+}~\left[\exists_{a, b \in \mathbb{N}^+}~\left(\left(q = \frac{a}{b}\right) \land \left(gcd(a, b) = 1\right)\right)\right]$
\item $\forall_{x}~\left(\left(x \in \mathbb{Q}\right) \in \mathbb{B}\right)$
\item $0 \in \mathbb{Q}$
\item $\mathbb{N}^+ \subset \mathbb{R}^+$
\item $\mathbb{Q} \subset \mathbb{R}$
\item $\mathbb{R}^+ \subset \mathbb{R}$
\item $1 < 2$
\item $1 \in \mathbb{N}^+$
\item $2 \in \mathbb{N}^+$
\item $\forall_{a, b}~\left(|\left(a, b\right)| = |\left(1, \ldots, 2\right)|\right)$
\item $\forall_{a, b \in \mathbb{R}~|~a > b}~\left(a \neq b\right)$
\item $\forall_{a \in \mathbb{R}^+}~\left(a > 0\right)$
\end{itemize}

\noindent Axioms required (directly or indirectly) to prove \\
\texttt{sqrt2\_is\_not\_rational}
\begin{itemize}
\item $\left(\top \land \top\right) = \top$
\item $\forall_{A~|~A = \top}~A$
\item $\forall_{A~|~A}~\left(A = \top\right)$
\item $\forall_{A \in \mathbb{B}~|~(\lnot A) \Rightarrow \bot}~A$
\item $\forall_{A \in \mathbb{B}~|~A \Rightarrow \bot}~(\lnot A)$
\item $(\lnot \bot) = \top$
\item $(\lnot \top) = \bot$
\item $\forall_{n \in \mathbb{N}^+}~\left[\forall_{P, Q}~\left(\begin{array}{c} \left[\exists_{x_{1}, \ldots, x_{n}~|~Q(x_{1}, \ldots, x_{n})}~P(x_{1}, \ldots, x_{n})\right] =  \\ 
    \lnot 
    \left[
    \begin{array}{l}
    \forall_{y_{1}, \ldots, y_{n}~|~Q(y_{1}, \ldots, y_{n})}\\
    \left(P(y_{1}, \ldots, y_{n}) \neq \top\right)
    \end{array}
    \right] \end{array}\right)\right]$
\item $\forall_{x, y}~\left(\left(x = y\right) \in \mathbb{B}\right)$
\item $\forall_{x, y}~\left(\left(y = x\right) = \left(x = y\right)\right)$
\item $\forall_{x, y, z~|~x = y, y = z}~\left(x = z\right)$
\item $\forall_{x, y}~\left(\left(x \neq y\right) = (\lnot \left(x = y\right))\right)$
\item $\forall_{f, x, y~|~x = y}~\left(f(x) = f(y)\right)$
\item $\forall_{A, B}~\left(\left(A \subset B\right) = \left(\left(A \subseteq B\right) \land \left(B \ncong A\right)\right)\right)$
\item $\forall_{A, B}~\left(\left(A \subseteq B\right) = \left[\forall_{x \in A}~\left(x \in B\right)\right]\right)$
\item $\forall_{x, S}~\left(\left(x \notin S\right) = (\lnot \left(x \in S\right))\right)$
\item $\forall_{n \in \mathbb{N}}~\left(\left(n + 1\right) \in \mathbb{N}\right)$
\item $0 \in \mathbb{N}$
\item $1 = \left(0 + 1\right)$
\item $2 = \left(1 + 1\right)$
\item $\forall_{x, y}~\left(\left(y > x\right) = \left(x < y\right)\right)$
\end{itemize}

%% file: russells_paradox.tex
What is a set?  What are sets allowed to contain?  If the definition is too broad, one is vulnerable to Russell's paradox~\cite{Irvine_Deutsch:2020_Russells_Paradox_EncyclopediaEntry}.  Consider, as Bertrand Russell did, defining the set of all sets that do not contain themselves.  Call this set $R$.  Now ask, is $R\in R$ ?  If we assume $R \in R$, then we can prove that $R \notin R$ since $R$ does not contain any set that contains itself.  If we assume $R \notin R$, then we can prove that $R \in R$ since $R$ contains all sets that do not contain themselves.  Thus, we cannot define $R$ to be a set (in which membership is clearly defined) without generating a contradiction.

There are multiple ways to define sets and avoid Russell's paradox.  In choosing our axioms [Appendix~\ref{sec:Axioms}], we essentially follow the standard approach of ZFC set theory~\cite{Kunen:1980_Set_Theory}.  First, we only allow a restricted form of set comprehension, known as specification or separation in set theory literature~\cite{Jech:2003_Set_Theory}.  Unrestricted set comprehension allows one to define sets according to any criterion for membership (e.g., ``Let $S$ be the set of all $x$ such that $Q(x)$ is true.'').  Russell's set is defined in this manner.  Restricted set comprehension only allows us to define subsets of existing sets in this manner (e.g. ``Let $S \subseteq T$ be the set of all $x \in T$ such that $Q(x)$ is true.'').  Second, all sets have a well-defined rank that is an ordinal number \cite{Rubin:1967_Set_Theory}.  Loosely speaking, the rank of a set is the depth to which sets are nested; the empty set has rank zero, and a set containing a set containing the empty set has a minimum rank of 2.  A set that contains itself cannot have a definite rank (the rank, $r$, would have to be larger than $r$ which is not possible for an ordinal number).  Furthermore, a set cannot contain a superset of itself, either directly or indirectly (as a member of a member, etc.) for the same reason.  In other words, the universe of sets (which is \emph{not}, itself, a set), is the von Neumann universe, also known as the cumulative hierarchy of sets~\cite{Jech:2003_Set_Theory}. \\

{\footnotesize
\begin{mdframed}[hidealllines=true, backgroundcolor=gray!10]

\vspace{0.1in}

\begin{center}\textbf{Box 1:\\ Avoiding Russel's paradox with irreflexive set membership}
\end{center}

Define Russell’s set with restricted set comprehension in the following way:  given any set $T$, we define $R_T = \{S \in T \;\rvert\; S \notin S\}$.   There are two conditions that any given $S$ must satisfy in order to belong to Russell's set, $R_T$.  First, $S$ must be an element of $T$.  Second, $S$ must satisfy the condition that $S \notin S$.

Define a set $T$ to be ``easy'' if:
\begin{enumerate}
    \item $\forall_{S \in T} \left[S \notin S\right]$.
    \item $\forall_{S \subseteq T} \left[S \notin S\right]$.
    \item $T \notin T$ (redundant from 2).
\end{enumerate}

We can prove that, if $T$ is easy, then $R_T = T$ (\textit{i.e.}, all members of $R_T$ are members of $T$ and vice-versa).

For all $S \in T$, $S \in R_T$ iff $S \notin S$.  $R_T$ is a subset of $T$.  Define the set $S^{\prime} = R_T$ and we assume for now that $S^{\prime}$ is eligible to be in $R_T$.  Then $(S^{\prime} \in R_T)$ iff $(S^{\prime} \notin S^{\prime})$. Therefore,  $(S^{\prime} \in R_T)$ iff $(S^{\prime} \notin R_T)$, and finally $(R_T \in R_T)$ iff $(R_T \notin R_T)$. This is a contradiction.

Assume that set membership is never reflexive (a set cannot contain itself).  Then all sets are ``easy'' and we can show that $S^{\prime}$ is not eligible to be in $R_T$ so we don't get a contradiction. Since $T$ is easy, for any $S \in T$ we have the true statement $S \notin S$.  Therefore $R_T = T$.   But $T \notin T$ because $T$ is easy.  Hence $S^{\prime} = R_T = T$ is not eligible to be in $R_T$ so there is no Russell contradiction.

\vspace{0.1in}

\end{mdframed}
}

Prohibiting unrestricted set comprehension and ensuring that membership is irreflexive (that is, no set may be a member of itself) will prevent Russell's paradox. Using restricted set comprehension, one may only define Russell's sets as parameterized by a superset $T$: $R_T = \{x~\in T~|~x \notin x\}$.  In this notation, $R_T$ is the subset of all elements of $T$ which do not contain themselves.  If sets are not allowed to contain themselves (\textit{i.e.}, set membership is irreflexive), $R_T = T$.\footnote{More precisely, without making implicit assumptions about equivalent sets being equal, $R_T \cong T$, meaning that $\forall_x (x \in R_T) = (x \in T)$.}  This irreflexive property will be maintained if we assume that $T$, and all of its members, are themselves irreflexive with respect to membership. (See Box 1 for some more detail.)

Ensuring membership is irreflexive (along with restricting set comprehension) is sufficient but not necessary to avoid Russell's paradox (for example, it is possible to allow sets that only contain themselves, known as Quine atoms~\cite{Forster:2003_Logic_Induction_Sets}).  
Ensuring that sets have a definitive, ordinal rank is sufficient but may not be necessary to ensure that membership is irreflexive.  The rank property, however, is useful for making meta-logical arguments regarding the consistency of axioms.  Using an inductive argument to prove that a given property is maintained by the axioms of a theory, the rank property is a more useful (stronger) induction hypothesis than the irreflexive membership property.  In any case, ZFC is a standard foundation of mathematics, so it is wise to follow its conventions for defining sets.  At a meta-logical level for determining and analyzing the axioms of \ProveIt{}, we therefore choose our sets to be defined to have a definitive, ordinal rank.\footnote{Our axioms may, however, be agnostic (incomplete) with respect to defining the rank of certain sets.  Unlike ZFC, our sets may contain elements that are not known to be sets and therefore have an unknown rank.  That is fine as long as the axioms are consistent with the existence of a well-defined rank for each set which ensures that sets do not directly or indirectly contain themselves or supersets of themselves.}

The main difference between our approach and ZFC set theory is that we never implicitly presume domains of discourse as we discussed in \S\ref{Sec:DomainsOfDiscourse}.  ZFC set theory presumes a domain of discourse of sets.  Every variable in ZFC represents a set.  In contrast, we use conditions that directly or indirectly involve set membership/non-membership when we want to restrict a corresponding domain to that of sets.  As long as our axioms only define membership/non-membership to be true for ``valid'' sets,\footnote{Here, ``valid set'' is a meta-logical concept meaning that the set has a definite, ordinal rank.} proper restrictions will be imposed with such conditions.  Importantly, no axiom states that $(x \in S)$ is a Boolean in general for any $S$.  It is only defined to be a Boolean if $S$ is a ``valid'' set, or is equivalent to a ``valid'' set with respect to set equivalence.

As an example, consider the following judgment that can be derived from our restricted comprehension axiom (axiom 1 of \texttt{proveit.logic.sets.comprehension} in Appendix~\ref{sec:Axioms}):
\begin{align}
    \vdash \forall_{S, Q, x}~
\left(\begin{array}{c} \left(x \in \left\{y~|~Q(y)\right\}_{y \in S}\right) \\  = 
\left[ \exists_{y \in S~|~Q(y)}~x = y \right]
\end{array}\right).
\end{align}
We may use this to define any set of the form 
$\left\{y~|~Q(y)\right\}_{y \in S}$ as the subset of $S$ that satisfies the condition defined by $Q$.  There is no explicit restriction on $S$; however, if this is instantiated with any $S$ which is not equivalent to  a ``valid set,'' it will not be possible to prove whether or not $x \in S$ is true for any $x$.  Therefore, you will not be able to prove that anything is or is not contained in your gibberish ``set.''  If one attempted to create a Russell-type set, $R_T$, with a $T$ that is not a proper set, one would \emph{not} be able to prove that anything is contained in $R_T$.  It would be consistent with assuming $T \cong R_T \cong \emptyset$.  Therefore, no contradiction would be possible.  Furthermore, while there is no explicit restriction on $Q$, its role in the condition of the existential quantifier ensures that we may interpret (at a meta-logical level) $Q(y)$ as $Q(y)=\top$.  Thus, while there is no restriction to ensure that $Q$ be instantiated with a function whose domain is $S$ and codomain is $\mathbb{B}$, it is only relevant that it map elements of $S$ to objects that are equal to true or not equal to true (which may include anything since we assume that every expression is either equal to true or not equal to true, though it is not always knowable which one it is).

We are not claiming to have a rigorous proof that we avoid Russell's paradox at this time.  We believe that we do avoid it (and that \ProveIt{}, with its basic axioms, is more broadly consistent), but there are tricky details that would be involved with such a proof.  Not only do we impose set domain restrictions indirectly through membership definitions, we also do not have explicit constraints on functions (they need not even be proper ``functions'' in the set-theoretic sense of having a specific domain and co-domain).  We will discuss this in the next section about Curry's paradox.

%% file: currys_paradox.tex
What is a function?  A definition that is too broad leaves one vulnerable to Curry's paradox.\footnote{Curry's paradox was discovered by Haskell Curry when he attempted to extend combinatory logic with logical connectives in order to include the extensionality axiom: two combinator expressions $e_1$ and $e_2$ are equivalent iff for every $x$, $e_1(x) = e_2(x)$. Kleene and Rosser \cite{Kleene_Rosser:1935_Inconsistency_of_Certain_Formal_logics} discovered a similar paradox in Church's extended system that 
also combined untyped lambda calculus with logical connectives using Richard's paradox (see an excellent article by Rosser \cite{Rosser:1982_History_of_Lambda_Calculus} explaining the
background of the paradox).  Curry came up with a cleaner version using Russell's paradox \cite{Curry:1941_paradox_of_kleene_and_rosser, Curry:1942_Inconsistency_of_Certain_Formal_Logics}.}
  The standard definition of a mathematical function (coming from ZFC set theory) is that it is a relation (ordered pair) between two sets (a domain and a codomain) in which there is a unique output (in the codomain) for every input (in the domain)~\cite{Kunen:1980_Set_Theory}.  This definition is considered to be ``safe,'' but \ProveIt{} does, in general, relax the requirement with respect to specifying a set-theoretic domain and codomain.  Such restrictions can be made [\textit{e.g.}, 
  $f \in \textrm{Map}(\mathbb{A}, \mathbb{B})$ could be defined as $\forall_{x \in \mathbb{A}}~f(x) \in \mathbb{B}$, effectively imposing $\mathbb{A}$ as the domain of $f$ and $\mathbb{B}$ as its codomain].  However, we have chosen some of our axioms to allow nonrestrictive ``functions.''   For example, the substitution axiom [see axioms  of \texttt{proveit.logic.equality} in Appendix~\ref{sec:Axioms}],
\begin{equation}
\vdash \forall_{f, x, y~|~x = y}~\left(f(x) = f(y)\right),
\end{equation}
allows us to substitute (replace) any $x$ for any $y$ in \emph{any} expression provided that $x=y$, regardless of whether or not $f$ has a domain containing $x$ and $y$.

Russell's paradox led to a contradiction by defining a set in a self-referential manner.  Applying a function in a self-referential manner can also lead to a paradox, as we will see by using Curry's paradox.  The paradox can be expressed in various logic systems~\cite{Shapiro_Beall:2018_Currys_Paradox} without proper restrictions in place.  Most relevant for \ProveIt{}'s purpose are the forms of the paradox in untyped lambda calculus and combinatory logic where it relates to the notion of a ``function.''  But let us start with the informal version that is easiest to explain.  Begin with a sentence of the form ``If this sentence is true, then the moon is made of cheese'' (analogous to examples in~\cite{Shapiro_Beall:2018_Currys_Paradox}).  Using ``reasonable'' arguments, we can derive ``the moon is made of cheese'' even though this is untrue.  Now use the law of deduction (if we can prove $B$ assuming $A$, then $A \Rightarrow B$).  In this case, we take ``this sentence is true'' as the assumption $A$ and use it to prove ``the moon is made of cheese'' as the consequent $B$ [since $A = (A \Rightarrow B)$ by definition].  In so doing, we have proven $A \Rightarrow B$ which is the original sentence.  Since $A$ represents our original sentence being true, we have thus proven $A$ and can then derive $B$ (for any $B$, true or not).  

In order to prove a contradiction using Curry's paradox, the analogue to ``this sentence is true'' must be constructed.  
We adapt a form of Curry's paradox from \cite{Rosser:1982_History_of_Lambda_Calculus} to illustrate how \ProveIt{} (we believe) avoids this pitfall.
Define $g$ as
\begin{equation}
P \mapsto (P(P) \Rightarrow \bot)
\end{equation}
using the notation of \ProveIt{}.\footnote{In classical untyped lambda calculus, $g = (\lambda P. (P(P) \Rightarrow A)$ and can be constructed using the fixed point $Y$ combinator since $g(g) = g$.}  Using beta reduction, it can be proved that
\begin{equation}
\vdash g(g) = (g(g) \Rightarrow \bot).
\end{equation}
Note that this has the same form as $A = (A \Rightarrow B)$.
With a few derivation steps, this will lead to a contradiction.  It is easy to see that $g(g) = (g(g) \Rightarrow \bot)$ is an equation with no solution.  $g(g)$ is either equal to true or not (as are all implications in \ProveIt{}).  If it is not equal to true, we have a contradiction because the right side will be true (in \ProveIt{}'s logic, an implication is always true when the antecedent is not true).  But if it is equal to true, we can prove $\top \Rightarrow \bot$ and thus $\bot$ (a contradiction).  This is similar to having an equation such as $x = x + 1$ in which there is no solution.

If we could create the expression of $g$ applied to itself without introducing an axiom, we would be in trouble.  That is, if we could express \\
$[P \mapsto (P(P) \Rightarrow A)](P \mapsto (P(P) \Rightarrow A))$
as an \exprtype{Operation} expression where the \subexpr{operator} and \subexpr{operand} are the same \exprtype{Lambda} expression, we'd be able to create an expression that is intrinsically paradoxical.\footnote{The analogue to this (a lambda expression as an operator) in lambda calculus is lambda application with lambda abstraction as the first term.  This is allowed in lambda calculus systems but not in \ProveIt{}.  Untyped lambda calculus, when extended with logical connectives, can be vulnerable to Curry's paradox for this reason.}  This is like Russell's paradox in which the expression $\{x~|~x \notin x\}$ is intrinsically paradoxical in na\"ive set theory.  Fortunately, our axioms do not admit an actual definition for $\{x~|~x \notin x\}$ (as we saw in Sec.~\ref{Sec:RussellsParadox}).  Similarly, our expression syntax does not allow $[P \mapsto (P(P) \Rightarrow A)](P \mapsto (P(P) \Rightarrow A))$ to be constructed.\footnote{For this same reason, the fixed-point ($Y$) combinator of combinatory logic cannot be constructed either.}

It is possible, in \ProveIt{}, to create $g(g)$ as an \exprtype{Operation} expression with $g$ as a \exprtype{Variable}.  We could then instantiate $g : P \mapsto (P(P) \Rightarrow \bot)$.  However, \ProveIt{}, according to the reduction rules described in Appendix~\ref{sec:ReductionRules}, will then greedily apply beta reduction until there is no more beta reduction to be performed.  In this case, however, that process never terminates.  So, for practical purposes, $g(g)$ with $g : P \mapsto (P(P) \Rightarrow \bot)$ cannot be expressed in \ProveIt{} without including an assumption or adding an erroneous axiom that defines $g = [P \mapsto (P(P) \Rightarrow \bot)]$.

At a meta-logical level, we can regard \ProveIt{} functions to have an implicit domain restriction that disallows any input that results in non-terminating beta reductions.  A function should be defined to have a single, definite evaluation given any particular input.  When beta reductions fail to terminate, it suggests that there is no single, definitive evaluation [e.g., $g(g) = (g(g) \Rightarrow \bot)$ has no definitive evaluation as we saw].  With this implicit domain restriction imposed on all functions, enforced by \ProveIt{}'s core rules and consistent with our basic axioms, we believe that Curry's paradox is avoided.  We admit that an implicit restriction of the domain of a function according to whether or not the reduction of the function application terminates is a nonstandard approach.  But it does appear to be effective, at least in regards to avoiding Curry's paradox.

%% file: core_types_and_literals.tex
\ProveIt{} uses an object-oriented framework for defining different types (classes) of expressions.  Each expression class defines how the expression is to be formatted (into \LaTeX{} or as a character string) and provides convenient methods for applying axioms or theorems/conjectures pertinent to the particular expression.
Each expression class must derive from on of the following primitive expression types that are defined in the core of \ProveIt{} (the \verb#proveit._core_# package):
\begin{description}
\item[Variable]
A label that is interchangeable (as long as it is kept distinct from different labels) with no intrinsic meaning.  It is often represented by a single letter ($a$, $b$, $x$, $y$, etc.) but can have any representation.
\item[Literal]
A label that is not interchangeable and has an intrinsic meaning.  Specific operators ($\lnot, \land, +, \times$, etc.) and specific irreducible values ($\top, \bot, 0, 5$, etc.) are all \exprtype{Literal}s.  Furthermore, a problem-story ``variable'' in a particular theory package, representing some unknown but particular value, should also be a \exprtype{Literal} (e.g., ``Ann has $a$ apples...'').
\item[ExprTuple]
Represents a finite, ordered list (sequence) of elements that may be used when there are multiple \subexpr{operands} of an \exprtype{Operation} (see next), or multiple \subexpr{parameters} of a \exprtype{Lambda} (see below).  An \exprtype{ExprTuple} contains zero or more \subexpr{entries} as sub-expressions.  An entry may represent a single element of the mathematical tuple being represented, or it may be an \exprtype{ExprRange} (see below) that represents any number of mathematical elements.  For example, \\
$(a, b, c_1, ... c_n)$ \\
has three \subexpr{entries} with the third entry being an \exprtype{ExprRange} that represents $n$ elements of the mathematical tuple.
\item[Operation]
The application of an \subexpr{operator} on \subexpr{operand(s)}.  For example, $0 + 5 + 8$, $A \lor B \lor C \lor D$, and $x < y$ are examples of \subexpr{operation} expressions.  The \subexpr{operator} must be a \exprtype{Literal}, \exprtype{Variable}, or an \exprtype{IndexedVar}.  They may \emph{not} be \exprtype{Lambda} expressions.  This is an important restriction that prevents Curry's paradox (discussed in Sec.~\ref{Sec:CurrysParadox}).
The \ProveIt{} theory packages define many types derived from the \exprtype{Operation} type (e.g., for each specific operation), but all operations have the same behavior with respect to reduction rules that will be discussed in Appendix~\ref{sec:ReductionRules}.
\item[Conditional]
An expression that has one \subexpr{condition} and one \subexpr{value}.  The \exprtype{Conditional} axiomatically reduces (equates) to its \subexpr{value} if and only if the \subexpr{value} is true.  For example $\{P(x) \textrm{ if } Q(x).$ reduces to $P(x)$ if $Q(x)$ is true.  Otherwise, the \exprtype{Conditional} is not defined (i.e., it is not reducible).  Furthermore, we have the following axiom \\
$\vdash \forall_{a, Q}~\left(\begin{array}{c} \left\{a \textrm{ if } Q\right.. \\  = \left\{a \textrm{ if } Q = \top\right.. \end{array}\right)$ \\
which defines the \exprtype{Conditional} only up the the truth of its \subexpr{condition} with no regard, beyond this, to whether or not the \subexpr{condition} has a Boolean value.  This is consistent with the manner in which \ProveIt{} judgments are treated without type restrictions.
\item[Lambda]
A mapping of \subexpr{parameter(s)} into a \subexpr{body} that may involve any or all of the \subexpr{parameters}.  For example, \\
$(x, y, z) \mapsto \{x+y/z~\textrm{ if }~x, y, z \in \mathbb{R}, z \neq 0.$ \\
is represented by a \exprtype{Lambda} with a \exprtype{Conditional} \subexpr{body} that converts three real numbers $x, y, z$ to $x + y/z$ as long as $z$ is not zero.  Each parameter represented by the \subexpr{parameter(s)} must be a \exprtype{Variable} or \exprtype{IndexedVar} (discussed below).  The \exprtype{Lambda} introduces a new scope for the \subexpr{parameters}.  The \subexpr{parameters} are said to be bound in this new scope; occurrences outside this scope are not deemed to be the same thing.  \exprtype{Lambda} expressions that are equivalent up to a relabeling of the parameters are deemed by \ProveIt{} to be the \emph{same} expression (or sub-expression); for example, \\
$(a, b, c) \mapsto \{a+b/c~\textrm{ if }~a, b, c \in \mathbb{R}, c \neq 0.$ \\
is deemed to be the same expression as the one above, simply using $a, b, c$ instead of $x, y, z$.  This is known as alpha conversion in the lambda calculus terminology and is discussed in Appendix~\ref{Sec:Relabeling} in more detail.
\item[NamedExprs]
A mapping from keyword strings to expressions.  This can be used to prevent ambiguity of an expression's internal representation when the order of a sequence of sub-expressions is not enough to specify the role of the sub-expression.
\item[ExprRange]
Represents a range of expressions with a \subexpr{parameter} going from a \subexpr{start\_index} to an \subexpr{end\_index} in successive unit increments ($+1$).  It contains a \exprtype{Lambda} sub-expression (\subexpr{lambda\_map}) to define each expression in the range as a function of the \subexpr{parameter} value (the index).
For example, $1/(x+i) + ... + 1/(x+j)$ is represented by an \exprtype{Operation} with `+' as the \subexpr{operator} and an \exprtype{ExprRange} as the \subexpr{operand}.  The \exprtype{ExprRange} sub-expression is $1/(x+i), ..., 1/(x+j)$.  The sub-expressions of this exprtype{ExprRange} may be denoted as \\
lambda\_map : $k \mapsto 1/(x+k)$ \\
start\_index : $i$ \\
end\_index : $j$
\item[IndexedVar]
A special kind of \exprtype{Operation} that indexes a \exprtype{Variable} with one or more indices.  It treats the variable being indexed as the \subexpr{operator} and the \subexpr{index} or \subexpr{indices} as the \subexpr{operand(s)}.  For example: $a_1$, $x_i$, $x_{i+1}$, $x_{i, j}$, etc.  An \exprtype{ExprRange} of \exprtype{IndexedVar}s acts as a collection of variables and may be used in an \exprtype{ExprTuple} of \exprtype{Lambda} \subexpr{parameters}.  For example:\\
$(a_1, \ldots, a_n, b_{1,1}, \ldots, b_{1,k}, \ldots\ldots, b_{j,1}, \ldots, b_{j,k})$ forms a valid collection of \exprtype{Lambda} \subexpr{parameters} with the $b$ variables being doubly indexed and contained in doubly-nested \exprtype{ExprRange}s.
\end{description}

These primitive expression types are special for the following reason.  In order to rigorously verify the correctness of a proof in \ProveIt{}, one must know
the DAG structure of the expressions involved with respect these primitive types.  Additionally, $\forall$ (forall), $\Rightarrow$ (implies), $\land$ (conjunction), and $=$ (equals) \exprtype{Literal}s play specific roles in the derivation and reduction rules.  However,
the derived expression classes are not important in this verification process (e.g., one need only know that $x+y$ is an \exprtype{Operation} with $+$ as a \exprtype{Literal} operator and $x$ and $y$ as \exprtype{Variable} operands, not that $x+y$ is an \texttt{Add} type expression).  
This is important for the sake of having a relatively small and stable \emph{core derivation system} for proof verification.

Classes are derived from these primitives for the purpose of defining specific ways of formatting expressions (as a string or as \LaTeX{}) and to provide methods for convenience and automation for manipulating expressions of that specific type, but this is external to the \emph{core derivation system} for proof verification.  The most common primitive type used as a base class is the \exprtype{Operation}; there are a wide variety of kinds of operations (logical connective operations such as $\land$, $\lor$, $\lnot$, etc.; the $\forall$ and $\exists$ quantifiers, number operations such as $+$, $-$, etc.).  Each one of these has its own methods that are convenient for applying axioms and theorems specific to the operation.

A few non-primitive expression classes are defined in the core (\verb#proveit._core_#) for organizational purposes.  These are:
\begin{description}

\item[OperationOverInstances]
An \exprtype{Operation} whose \subexpr{operand} is a \exprtype{Lambda} expression.  Often, the \exprtype{Lambda} expression has a \exprtype{Conditional} body.  The idea is that it operates over the domain of instances of the \exprtype{Lambda} parameter variables for which the \subexpr{condition} of the \exprtype{Conditional} is satisfied.  Quantifiers are \exprtype{OperationOverInstances}.  $\forall_{x, y~|~Q(x, y)} P(x, y)$ is internally represented by \ProveIt{} as an operation of the $\forall$ operator acting on $(x, y) \mapsto \left\{P(x, y) \textrm{ if } Q(x, y)\right.$.

\item[ConditionalSet]
An \exprtype{Operation} that operates on any number of \exprtype{Conditional} expressions.  If one and only one of the \subexpr{condition} expressions is satisfied (proven true), and the rest are proven to be untrue (though not necessarily false), it axiomatically reduces (equates) to the \subexpr{value} associated with the satisfied \subexpr{condition}.  For example, consider this definition for the absolute value of an integer: \\
$\left\{
\begin{array}{l}
    x ~\textrm{ if }~x \geq 0 \\
    -x ~\textrm{ if }~x < 0
\end{array}
\right..$ \\
This would be represented by a \exprtype{ConditionalSet} with two \exprtype{Conditional} \subexpr{operands}.
\item[ExprArray]
An \exprtype{ExprTuple} of \exprtype{ExprTuple}s formatted as a 2-dimensional array.  It may contain \exprtype{ExprRange}s.  For example, \\
$\begin{array} {cccccc} 
 A_{i} & \cdots & A_{j} & B_{i} & \cdots & B_{j}  \\ 
  C_{i} & \cdots & C_{j} & D_{i} & \cdots & D_{j} 
\end{array}$ \\
is an \exprtype{ExprArray} that is a tuple of two tuples, each containing ranges of indexed variables, formatted in a ``horizontal'' layout style.

\end{description}

%% file: styles.tex
As far as the \ProveIt{} software is concerned, the meaning of an expression is entirely dictated by the expression DAG with respect to the primitive base classes.  The actual formatting of an expression (in \LaTeX{} or as a text string) is a separate matter and there are no safeguards to ensure that the formatted expression reflects its meaning.  Of course, in developing \ProveIt{} we do our best to make sure that formatting routines will make the notation clear and unambiguous.  But by not placing restrictions on the formatting aspect, we have a lot of freedom in rendering elegant mathematical representations in \LaTeX{}, and can offer flexibility in the style choices of this formatting.  To be certain about the genuine meaning of an axiom or a theorem, users are encouraged to view the full expression DAG.  The elegant \LaTeX{} representations, however, are extremely convenient to aid one's intuition about the mathematical constructs being utilized.

\subsection{Style options}
\label{Sec:StyleOptions}

Each expression class has its own style options which can be modified to change the formatting of an expression without changing its meaning.  For example, \\
$a / b$ and $\frac{a}{b}$ \\
are both manifestations of the same expression of the \verb#Div# class defined in the \verb#proveit.numbers.division# theory package.  The first one has its \verb#division# style set to \verb#inline#; the second one has its \verb#division# style set to \verb#fraction#.  
A few other examples were already mentioned in Appendix~\ref{Sec:ExprTypes}.

Two expressions that are the same apart from their style are deemed to be the \emph{same} expression in \ProveIt{}.
The python \verb#==# operation will return \verb#True# when applied to two expressions that are the same up to their style settings.  They are not the same object in every respect, of course, if they have different style settings.  But they are the same with respect to proof derivations which is the entire purpose of \ProveIt{}.

Style options may be used as an alternative to defining new expression classes and associated axioms for defining new notation.  Table \ref{tab:style_alternates} lists many examples.  Whether something should be done as a style option or as a separate expression type is not always obvious.  There are trade-offs that must be balanced when deciding whether to define notation using style or a new expression class.  Regarding $a < b \leq c < d$ as a shorthand notation for $(a < b) \land (b \leq c) \land (c < d)$ is very sensible; it simplifies the code and proofs, and is relatively straightforward for the user. As a counterexample, it is very tempting to regard $x \neq y$ as shorthand for $\lnot(x = y)$.  However, by defining a separate \texttt{NotEquals} class, we are able to write convenient and accessible methods for manipulating this specific kind of operation independent of the negation operation methods.

\begin{table}
\begin{tabular}{c|c}
\hline
    $\forall_{x~|~Q(x) \land R(x)} P(x)$ &  $\forall_{x~|~Q(x), R(x)} P(x)$ \\
\hline
    $\{\}$ & $\emptyset$ \\
\hline
    $A \subseteq B$ & $B \supseteq A$ \\
\hline
    $A \subset B$ & $B \supset A$ \\
\hline
    $(A \subseteq B) \land (B \subset C)$ & $A \subseteq B \subset C$ \\
\hline
    $x \leq y$ & $y \geq x$ \\
\hline
    $x < y$ & $y > x$ \\
\hline
    $(x < y) \land (y \leq z)$ & $x < y \leq z$ \\
\hline
    $w + (-x) + y + (-z)$ & $w - x + y - z$ \\
\hline
\end{tabular}
\caption{Examples of alternate style options. The left side best reflects the internal representation.  Variables are arbitrary just for examples.  Line wrapping and choices regarding parentheses should also by done as style options (though not all of these are supported yet).} 
\label{tab:style_alternates}
\end{table}

\subsection{Parameter labels as a style choice}
\label{Sec:Relabeling}
As mentioned in Appendix~\ref{Sec:ExprTypes} in the definition of the \exprtype{Lambda} expression class,  
\exprtype{Lambda} expressions that are equivalent up to a relabeling of the parameters are deemed by \ProveIt{} to be the \emph{same} expression.  While the machinery for this is different than the machinery for setting style options described above, it is essentially a style feature.  Changing   \exprtype{Lambda} parameters does not alter the meaning of the expression, it only changes the way in which it is formatted.  In lambda calculus terminology, this is known as alpha conversion.

\ProveIt{} implements this feature by generating a canonical form whenever a \exprtype{Lambda} expression is created.  This is in the same spirit as De Bruijn notation used in lambda calculus though it does work differently (De Bruijn notation may assign different indices to different occurrences of the same variable, while we assign a consistent label for all occurrences of a variable in the same scope).  Essentially, our canonical form assigns parameters to the first unused dummy variable going from $_{-}a$ to $_{-}z$ and continuing $_{-}aa$ to $_{-}az$, $_{-}ba$ to $\_bz$, etc. (it is unlikely so many variables would be needed, but there are endless possible dummy variables).
For example, the canonical version of \\
$x \mapsto \left[\exists_{y}~\left(\left(x + y + z\right) = 0\right)\right]$ \\
is \\
${_{-}b} \mapsto \left[\exists_{{_{-}a}}~\left(\left({_{-}b} + {_{-}a} + z\right) = 0\right)\right].$

When there are a range of parameters, the canonical form is typically established in a similar manner.  The canonical form of \\
$\left(x_{1}, \ldots, x_{n}, y_{1}, \ldots, y_{n}\right) \mapsto \left(\left(x_{1} \cdot y_{1}\right) + \ldots + \left(x_{n} \cdot y_{n}\right)\right)$ \\
is \\
$\left({_{-}c}_{1}, \ldots, {_{-}c}_{n}, {_{-}b}_{1}, \ldots, {_{-}b}_{n}\right) \mapsto \\
\left(\left({_{-}c}_{1} \cdot {_{-}b}_{1}\right) + \ldots + \left({_{-}c}_{n} \cdot {_{-}b}_{n}\right)\right)
.$ \\
In this style it is not obvious why ${_{-}a}$ is skipped, but if we change the style option of the \exprtype{ExprRange} within the \verb#Add# \exprtype{Operation} so that the \verb#parameterization# is \verb#explicit#, we have \\
$\left({_{-}c}_{1}, \ldots, {_{-}c}_{n}, {_{-}b}_{1}, \ldots, {_{-}b}_{n}\right) \mapsto \\ \left(\left({_{-}c}_{1} \cdot {_{-}b}_{1}\right) + ..\left({_{-}c}_{{_{-}a}} \cdot {_{-}b}_{{_{-}a}}\right).. + \left({_{-}c}_{n} \cdot {_{-}b}_{n}\right)\right)$ \\
where we now see that ${_{-}a}$ is being used to parameterize the \exprtype{ExprRange} that runs from $1$ to $n$.

In the above example, there is no question that all of the utilized indices of $x$ and $y$ were covered by the parameters, $\left(x_{1}, \ldots, x_{n}, y_{1}, \ldots, y_{n}\right)$.  Therefore, we are free to relabel $x$ and $y$.  However, for expressions where there is any ambiguity (or is not sufficiently obvious) whether the parameters cover all utilized indices, relabeling for those variables is disabled.  For example, $x$ and $y$ may not be relabeled in \\
$\left(x_{1}, \ldots, x_{n}, y_{1}, \ldots, y_{n}\right) \mapsto \left(\left(x_{1} \cdot y_{1}\right) + \ldots + \left(x_{m} \cdot y_{m}\right)\right)$ \\
and therefore $x$ and $y$ remain as the labels in the canonical form.  This expression alone gives no indication as to whether or not $m > n$.  If $m > n$ in this example, there would be indices of $x$ and $y$ that are not covered by the ranges of parameters, and therefore there are free indexed variable occurrences (e.g., $x_{n+1}$ and $y_{n+1}$ are free).  It would therefore be improper to relabel $x$ and $y$.  Even in instances where the coverage may seem obvious, as in \\
$\left(x_{1}, \ldots, x_{n}\right) \mapsto \left(x_{1} + \ldots + x_{n-1} + x_n\right)$ \\
\ProveIt{} may error on the side of caution (and simplicity) and still regard $x$ as non-relabel-able.  The \infrule{Instantiation} inference rule [Appendix~\ref{Sec:Instantiation}] offers more versatility and can effect relabeling in instances where canonical relabeling is not allowed.  As an actual derivation step, it has more versatility than is allowable by a mere style feature (a derivation step can have dependencies to satisfy requirements while the justification of a style feature must be intrinsic to the expression itself).

%% file: inference_rules.tex
In this section, we describe each of \ProveIt{}'s small set of inference rules used to construct derivation steps of a proof.  We will use the standard notation for inferences rules with a horizontal line separating premises above the line from a conclusion below the line:
\[
    \begin{tabular}{l}
         Premises
         \\[-1pt]\hline
         \\[-10pt]
         Conclusion
    \end{tabular}
\]
The premises represent any number of judgments and the conclusion represents a single judgement that may be derived from the premises.  As we use this inference rule notation, we may either present explicit judgment assumptions within curly braces, or we will represent assumption sets in a set theoretic manner (e.g., $S \cup T \vdash X$ has, as assumptions the set union of $S$ and $T$).  When variable dependencies are relevant, the variable dependence will be explicitly shown.  For example, we will use $P(x)$ to
denote an expression in which $x$ may (or may not) appear as a free variable, and we use $T(x)$ to denote a set of assumptions for which $x$ may (or may not) appear as a free variable in each assumption. 

\subsection{Proof by assumption}
\label{Sec:Assumption}

  In \ProveIt{}, any expression may be proven to be true by assumption.  That is, \\
  \[
    \begin{tabular}{l}
      \\ [2pt]\hline
      \\[-10pt] 
      $\{A\} \vdash A$
    \end{tabular}\tag{ASSUMPTION}
\]
  may be introduced as a valid judgment in a proof for any $A$.  It does not matter if $A$ is an expression that would normally have a Boolean type, and one must never assume that it has an intrinsic Boolean type just because it may be assumed.  By assuming it, it is true (and therefore Boolean) \emph{under that assumption}.  To emphasize this point, note that we may prove, in \ProveIt{}, that \\
  $\{3\} \vdash 3$ \\
  which should simply be taken to mean that \\
  $\{3 = \top\} \vdash 3=\top$ \\
  as was discussed in Sec.~\ref{Sec:DomainsOfDiscourse}.
  Of course, this assumption is bogus, so whatever conclusion one deduces under this assumption is worthless (but correct by \ProveIt{}'s logic).
  
  One may also have a range of assumptions using an \exprtype{ExprRange}.  Using a range of assumption, we can prove that the conjunction of these expressions is true.  Thus,
    \[
    \begin{tabular}{l}
      \\ [2pt]\hline
      \\[-10pt] 
      $\{A_1, \ldots, A_m\} \vdash A_1 \land \ldots \land A_m$
    \end{tabular}\tag{ASSUMPTIONS}
\]
  Notice that we have no assumption here about $m$ being a natural number.  However, for this to be a valid assumption, $m$ must be a natural number.  But, again, we are not concerned about what we are able to prove under invalid assumptions.  Introducing a logical conjunction here is important.  It would never be valid (under valid assumptions) to prove that a tuple of expressions is true.  Rather, it's the conjunction of these expressions, in this case, that forms the statement we are claiming to be true.  In a sense, there is an implicit conjunction as a connective joining all assumptions.  But it is only when a range of assumptions appears on the right of the turnstile that we need to make it explicit.
  \hfill \\
  
\subsection{Axiom/theorem/conjecture invocation}
\label{Sec:ThmInvocation}
  
  Another way to introduce a judgment into a proof is by invoking an axiom, theorem, or even a conjecture.  An axiom is regarded to be true as an asserted fact.  Theorems are as valid as the axioms used, directly or indirectly, to prove them.  A conjecture in \ProveIt{} may be a well-known fact, but its conjecture status indicates that it has not be fully proven in the \ProveIt{} system; it either has no proof, or its proof relies, itself, on conjectures.  Within a given proof, however, these are accepted as valid judgments without question.  The dependencies of a proof, discussed in Sec.~\ref{Sec:TheoremDependencies}, provide the additional information about what axioms were required by the proof as well as what unproven conjectures were relied upon.
  
\subsection{Modus ponens}
\label{Sec:ModusPonens}

We can express our \infrule{Modus Ponens} inference rule, which is a generalization of standard modus ponens as

\[
    \begin{tabular}{l}
      $S \vdash A$ \hspace*{10pt} $T \vdash A \Rightarrow{}{} B$
      \\ [2pt]\hline
      \\[-10pt] 
      $S \cup T \vdash B$
    \end{tabular}\tag{MP}
\]
This reduces to the standard modus ponens rule when $S$ and $T$ are both the empty sets.  Modus ponens allows one to derive the consequent of an implication by proving its antecedent.

\subsection{Deduction}
\label{Sec:Deduction}
  
The inference rule of \infrule{Deduction} is:
\[
    \begin{tabular}{l}
  $S \cup \{A\} \vdash B$ 
  \\[2pt] \hline
  \\[-10pt] 
  $S \vdash A \Rightarrow{}{}B$
\end{tabular} \tag{D}
\]
which is essentially the inverse of modus ponens.  Modus ponens is used to eliminate an implication, deduction is used to introduce an implication.  As a consequence of our \infrule{Modus Ponens} and \infrule{Deduction} rules and the fact that $\{A\} \vdash B$ in \ProveIt{} means the same as $\{A=\top\} \vdash B=\top$, as discussed in Sec.~\ref{Sec:DomainsOfDiscourse},  $A \Rightarrow B$ means the same as $(A=\top) \Rightarrow (B=\top)$ in the \ProveIt{} system.

\subsection{Instantiation}
\label{Sec:Instantiation}

The \infrule{Instantiation} rule eliminates universal quantifiers, instantiating their variables to \emph{any} expression.  The following expresses the \infrule{Instantiation} rule when eliminating one universal quantifier for one variable:
\[
    \begin{tabular}{l}
         $S \vdash \forall_{x~|~Q(x)}~P(x)$ \hspace*{10pt} $T \vdash Q(Y)$
         \\[-1pt]\hline
         \\[-10pt]
         $S \cup T \vdash P(Y)$
    \end{tabular} 
\]
where $x$ is a variable that may (or may not) occur free in $Q(x)$ and $P(x)$, and $Y$ is \emph{any} expression and serves as the replacement for $x$.  There are specific reduction rules that dictate precisely how these replacements are to be performed that are described in Appendix \ref{sec:ReductionRules}.  In particular, this is implemented by effectively creating an ad-hoc lambda map for 
$x \mapsto Q(x)$ and $x \mapsto P(x)$ and applying these each to $Y$ (performing beta reduction as it is called in lambda calculus terminology) to produce $Q(Y)$ and $P(Y)$, possibly performing other reductions in the process.  These reduction rules may have requirements of their own which will be effectively added to the premises of the inference rule with their own assumption sets that must join the assumption set of the new judgment.  Furthermore, if $Q(x)$ is a conjunction of conditions, such as $Q_1(x) \land Q_2(x) \land Q_3(x)$, these will effectively be individual premises.  That is,
\[
    \begin{tabular}{l}
         $S \vdash \forall_{x~|~Q_1(x), Q_2(x), Q_3(x)}~P(x)$ \\
         $T_1 \vdash Q_1(Y)$ \hspace*{10pt} $T_2 \vdash Q_2(Y)$ \hspace*{10pt} $T_3 \vdash Q_3(Y)$
         \\[-1pt]\hline
         \\[-10pt]
         $S \cup T_1 \cup T_2 \cup T_3 \vdash P(Y)$
    \end{tabular} 
\]
Although $Q_1(x), Q_2(x), Q_3(x)$ show up as a comma delimited list of conditions of the universal quantifier, this is merely a style choice for depicting a conjunction of conditions. 

More generally, we may instantiate any number of variables:
\[
    \begin{tabular}{l}
         $S \vdash \forall_{a, b, \ldots, z~|~Q(a, b, \ldots, z)}~P(a, b, \ldots, z)$ \\
         $T \vdash Q(A, B, \dots, Z)$
         \\[-1pt]\hline
         \\[-10pt]
         $S \cup T \vdash P(A, B, \dots, Z)$
    \end{tabular} \tag{INST}
\]
where $a, b, \ldots, z$ is used to represent one or more variables which may include ranges of indexed variables and ranges of ranges of indexed variables, etc. via the \exprtype{ExprRange} and \exprtype{IndexedVar} expression types.  $A, B, \dots, Z$ are their respective replacements.  Again, there are the caveats that a premise will be required for each \subexpr{entry} of a conjunction of conditions and there may be additional premises from requirements of the lambda applications and related reductions resulting from replacing $a, b, \ldots, z$  with $A, B, \dots, Z$ to produce $P(A, B, \ldots, Z)$ and $Q(A, B, \ldots, Z)$.

\ProveIt{}'s \infrule{Instantiation} rule may be used to eliminate any number of nested universal quantifiers for convenience as well as for the sake of brevity in the proof DAGs.  The rule is essentially the same as above if you treat the nested universal quantifiers as a single universal quantifier over the union of the parameters and joining all of the conditions into one flattened conjunction.  One may also regard this as multiple applications of the rule described here, but some relabeling of parameters along the way may be necessary for capture-avoidance purposes (discussed in Appendix~\ref{Appendix:CaptureAvoidance}).

\subsection{Generalization} 
\label{Sec:Generalization}  

The \infrule{Generalization} rule introduces an explicit universal quantification over any number of variables on the right side of the turnstile.  There is essentially an implicit universal quantification over free variables of judgments acknowledging that a free variable appearing on both sides of the turnstile is the \emph{same} variable.  That means, if we are to add an explicit universal quantification over, say, $x$, on the right side of the turnstile, any assumptions involving $x$ as free must be appropriately moved to the right side where it is now bound by the scope of the added quantifier.  This is appropriately done by making such assumptions serve as \subexpr{conditions} of the universal quantifier.  A relatively simple form of the \infrule{Generalization} rule is
\[
    \begin{tabular}{l}
         $S \cup T(x) \vdash P(x)$
         \\[-1pt]\hline
         \\[-10pt]
         $S \vdash \forall_{x~|~[\land](T(x))} P(x)$ 
    \end{tabular} 
\]
where $[\land](T(x))$ is meant to denote a conjunction (in any order) over the assumptions, $T(x)$, that may (or may not) involve $x$ as a free variable.  This is similar to the ASSUMPTIONS rule where we needed to introduce a conjunction when performing proof-by-assumption on a range of assumptions.  The assumptions of $S$ must not have $x$ as a free variable.

Adding extraneous conditions to the universal quantifier can only weaken the statement.  We therefore allow, more generally,
\[
    \begin{tabular}{l}
         $S \cup T(x) \vdash P(x)$
         \\[-1pt]\hline
         \\[-10pt]
         $S \vdash \forall_{x~|~[\land](Q(x), T(x))} P(x)$
    \end{tabular} 
\]
where $[\land](Q(x), T(x))$ is meant to denote a conjunction (in any order) over all of the assumptions, $T(x)$, that may (or may not) involve $x$ as a free variable and some extra conditions, $Q(x)$ (which may or may not involve $x$) that only weakens the statement.  More generally, we may add a universal quantifier for any number of variables:
\[
    \begin{tabular}{l}
         $S \cup T(a, b, \ldots, z) \vdash P(a, b, \ldots, z)$
         \\[-1pt]\hline
         \\[-10pt]
         $S \vdash \forall_{a, \ldots, z~|~[\land](Q(a, \ldots, z), T(a, \ldots, z))} P(a, \ldots, z)$
    \end{tabular} \tag{GEN}
\]
where $a, b, \ldots, z$ (as well as $a, \ldots, z$ for brevity) is used to represent one or more variables which may include ranges of indexed variables and ranges of ranges of indexed variables, etc. via the \exprtype{ExprRange} and \exprtype{IndexedVar} expression types.

\ProveIt{}'s \infrule{Generalization} rule may be used to introduce any number of nested universal quantifiers for the sake of brevity in the proof DAGs, but one may simply regard this as multiple applications of the rule described here.

Much like \infrule{Modus Ponens} and \infrule{Deduction} are essentially inverses of each other for eliminating and introducing an implication, \infrule{Instantiation} and \infrule{Generalization} are essentially inverses of each other for eliminating and introducing a universal quantifier.  As a consequence of our \infrule{Instantiation} and \infrule{Generalization} rules and the fact that $\{Q(x)\} \vdash P(x)$ is the same as $\{Q(x) = \top\} \vdash P(x) = \top$ in \ProveIt{}, $\forall_{x~|~Q(x)}~P(x)$ must mean the same as $\forall_{x~|~Q(x)=\top}~P(x)=\top$ in the \ProveIt{} system.

  \hfill \\
  \textbf{Literal Generalization (axiom elimination)} \hfill \\

The essential difference between a \exprtype{Variable} and a \exprtype{Literal} is that a \exprtype{Variable} is scoped within a particular judgment (and sometimes within a \exprtype{Lambda} expression within the judgment) whereas a \exprtype{Literal} has a universal scope (across judgments).  Consider a judgment $\vdash P(\mathfrak{a})$ involving some \exprtype{Literal}, $\mathfrak{a}$, that has been fully proven (to the level of axioms with no unproven conjectures).  Now collect all of the required axioms for this proof that involve $\mathfrak{a}$.  Say there are $k$ of them denoted as $\vdash A_1(\mathfrak{a})$, \ldots, $\vdash A_k(\mathfrak{a})$.  Then it must be the case that \\
$\{A_1(a), \ldots, A_k(a)\} \vdash P(a)$ \\
using all of the axioms before \emph{except} $\vdash A_1(\mathfrak{a})$, \ldots, $\vdash A_k(\mathfrak{a})$ (which are no longer necessary).  We have replaced the \exprtype{Literal} $\mathfrak{a}$ with the \exprtype{Variable} $a$ since we have brought all of the relevant facts pertaining to $\mathfrak{a}$ into the scope of a single judgment.

More generally, we can perform this transformation on any number of \exprtype{Literal}s simultaneously.  This is not a typical inference rule since the axiom requirements of the conclusion are reduced relative to the axiom requirements of the premise, but we shall denote this as
\[
    \begin{tabular}{l}
         $S(\mathfrak{a}, \mathfrak{b} \ldots, \mathfrak{z}) \vdash P(\mathfrak{a}, \mathfrak{b} \ldots, \mathfrak{z})$ \\
         \textrm{ via axioms } $\mathcal{A}(\mathfrak{a}, \mathfrak{b} \ldots, \mathfrak{z}) \cup \mathcal{B}$
         \\[-1pt]\hline
         \\[-10pt]
         $S(a, b \ldots, z) \cup \mathcal{A}(a, b \ldots, z) \vdash P(a, b \ldots, z)$ \\
         \textrm{ via axioms } $\mathcal{B}$
    \end{tabular} \tag{AXIOM\_ELIM}
\]
where $\mathfrak{a}, \mathfrak{b} \ldots, \mathfrak{z}$ are one or more \exprtype{Literal}s which are respectively replaced with \exprtype{Variable}s $a, b \ldots, z$.  $\mathcal{A}(\mathfrak{a}, \mathfrak{b} \ldots, \mathfrak{z})$ denotes the set of axiom expressions (the right side of axiom judgments which are always assumption-less by our convention) of required axioms that involve any of the \exprtype{Literal}s being converted, and $\mathcal{B}$ are the axiom expressions of required axioms that do not involve any of these \exprtype{Literal}s.

With \exprtype{Literal}s converted to \exprtype{Variable}s and axioms converted to assumptions, we may now generalize these and/or any other \exprtype{Variable}s using the GEN rule.  Combining AXIOM\_ELIM and GEN, we have a LITERAL\_GEN rule as a single derivation step in \ProveIt{}.  For example, if we have fully proven some $\vdash P(\mathfrak{a})$ via axioms $\vdash A(\mathfrak{a})$ (and no other axioms involving $\mathfrak{a}$, we may derive \\
$\vdash \forall_{a~|~A(a)} P(a)$ \\
in one LITERAL\_GEN step, eliminating the $\vdash A(\mathfrak{a})$ as requirements in the process.

%% file: reduction_rules.tex
As discussed in Appendix \ref{Sec:Instantiation}, when the \infrule{Instantiation} inference rule is invoked,
it creates ad-hoc \exprtype{Lambda} expressions that define functions and applies those functions to operands determined by the desired replacements of the instantiation.  The manifestation of the inference rule is dictated by these lambda application results.  For example, if we instantiate
\begin{align*}
    \vdash \forall_{x, y \in \mathbb{N}}~(x + y) \in \mathbb{N}
\end{align*}
with $x : 5$ and $y : b$ under the assumption that $b \in \mathbb{N}$, this results in
\begin{align*}
    \{b \in \mathbb{N}\} \vdash (5 + b) \in \mathbb{N}
\end{align*}
because $(x, y) \mapsto x \in \mathbb{N}$ applied to $(5, b)$ results in $5 \in \mathbb{N}$ (which is known to be true via axioms in the {\small\texttt{proveit.numbers}} theory package), $(x, y) \mapsto y \in \mathbb{N}$ applied to $(5, b)$ results in $b \in \mathbb{N}$ (which is true by assumption), and $(x, y) \mapsto (x + y) \in \mathbb{N}$ results in $(5 + b) \in \mathbb{N}$.

Lambda application is one reduction rule that \ProveIt{} uses during \infrule{Instantiation}.  Other reduction rules may be triggered in the process of performing a lambda application and making replacements of variables or ranges of variables as dictated by the lambda application.  Reduction rules are applied under a set of \emph{assumptions} and may generate their own requirements.  The assumptions come from the \infrule{Instantiation} invocation (i.e., the \verb#instantiate# method of the \verb#Judgment# class).   The \emph{requirements} will be effectively added to the premises of the inference rule of the instantiation and are allowed to use any of these assumptions.  The actual assumptions of the concluding judgment of the effected \infrule{Instantiation} will be the union of the assumptions actually used by the premises.  If an assumption is not utilized, it will be dropped.

\subsection{Lambda application (beta reduction)}
\label{Sec:Application}

The \reductionrule{lambda application} reduction rule (beta reduction using lambda calculus terminology) needs to match operands to parameters, ensuring lengths match in the case of parameter ranges (parameters involving \exprtype{ExprRange}s of \exprtype{IndexedVar}s), to establish a mapping between parameters (or ranges of parameters) and their replacements.  Then the replacements (substitutions) are performed, invoking other reduction rules in the process as appropriate.  

As a simple example,
\begin{align*}
    \left(x, y, z\right) \mapsto \left(\left(x + y\right) \cdot z\right)
\end{align*}
applied to $(a + x, b \cdot y, b + y + x)$ yields
\begin{align*}
    \left(\left(a + x\right) + \left(b \cdot y\right)\right) \cdot \left(b + y + x\right)
\end{align*}

When there are ranges of parameters, it is a little more interesting.  For example,
\begin{align*}
    \left(x_{1}, \ldots, x_{n}, y_{1}, \ldots, y_{n}\right) \mapsto \left(\left(x_{1} \cdot y_{1}\right) + \ldots + \left(x_{n} \cdot y_{n}\right)\right)
\end{align*}
applied to
\begin{align*}
    \left(1 \cdot 1, \ldots, n \cdot n, 1 + 1, \ldots, n + n\right)
\end{align*}
under the assumption $n \in \mathbb{N}$ yields
\begin{align*}
    \left(\left(1 \cdot 1\right) \cdot \left(1 + 1\right)\right) + \ldots + \left(\left(n \cdot n\right) \cdot \left(n + n\right)\right)
\end{align*}
with the requirements that
\begin{align*}
    \{n \in \mathbb{N}\} &\vdash |\left(1 \cdot 1, \ldots, n \cdot n\right)| = |\left(1, \ldots, n\right)|\\
    \{n \in \mathbb{N}\} &\vdash |\left(1 + 1, \ldots, n + n\right)| = |\left(1, \ldots, n\right)|
\end{align*}
to ensure that the number of \subexpr{operands} equals the corresponding number of \subexpr{parameters}.  Applying this same example lambda map to
\begin{align*}
    \left(0 \cdot 0, \ldots, k \cdot k, x, 1, \ldots, m, 0 + 0, \ldots, k + k, y, 1, \ldots, m\right)
\end{align*}
under the assumption that $k + 1 + m = n$ yields \\
$\left(\left(0 \cdot 0\right) \cdot \left(0 + 0\right)\right) + \ldots + \left(\left(k \cdot k\right) \cdot \left(k + k\right)\right) + \left(x \cdot y\right) + \left(1 \cdot 1\right) + \ldots + \left(m \cdot m\right)$ \\
with the requirements of \\
$\{k + 1 + m = n\} \vdash |\left(0 \cdot 0, \ldots, k \cdot k, x, 1, \ldots, m\right)| = |\left(1, \ldots, n\right)|$, \\
$\{k + 1 + m = n\} \vdash |\left(0 + 0, \ldots, k + k, y, 1, \ldots, m\right)| = |\left(1, \ldots, n\right)|$,
again ensuring that the number of \subexpr{operands} equals the corresponding number of \subexpr{parameters}.

A variable may occur in an expression in various forms, indexed over different ranges.  In order to treat the various forms that a range of parameters may occur in an unambiguous and versatile manner, \ProveIt{} allows one to specify ``equivalent alternative expansions''
for specifying various expansions for the different alternative forms.  The rule in doing this is fairly simple and straightforward, but allows for a lot of versatility.  Basically, if $x_i, ..., x_j$ is a range of parameters of the lambda expression,  $(x_i, x_{i+1}, ..., x_{j-1}, x_j)$ could be an equivalent alternative expansion of $(x_i, ..., x_j)$ assuming $j-i \geq 1$ and have have its own replacement.  These alternative expansions can provide the information needed to expand the variable in its various forms in \exprtype{ExprRange} expansions described below.  The requirements to allow for these alternative expansions are fairly straightforward.  The tuple of indices of the
equivalent alternative expansions must be equal to each other
(e.g., $(i, i+1, ..., j-1, j) = (i, ..., j)$), 
and their replacements must be equal to each other.  For example, \\
$\left(A_{1}, \ldots, A_{m}\right) \mapsto \\
\left(A_{1} \land \ldots \land A_{j} \land \left[\forall_{A_{i}, \ldots, A_{j}}~\left(A_{i} \lor \ldots \lor A_{j}\right)\right] \land A_{m}\right)$ \\
applied to
$\left(\lnot B_{1}, \ldots, \lnot B_{i - 1}, \lnot C_{1}, \ldots, \lnot C_{i}, A \lor D\right)$ \\
assuming $i \in \mathbb{N}^+$, $j = 2 \cdot i - 1$ and $m=j+1$ and using $\left(A_{1}, \ldots, A_{i - 1}, A_{i}, \ldots, A_{j}, A_{m}\right)$ and $\left(A_{1}, \ldots, A_{j}, A_{m}\right)$ as equivalent alternative expansions for $\left(A_{1}, \ldots, A_{m}\right)$ yields \\
$(\lnot B_{1}) \land \ldots \land (\lnot B_{i - 1}) \land (\lnot C_{1}) \land \ldots \land (\lnot C_{i}) \\
\land \left[\forall_{A_{i}, \ldots, A_{j}}~\left(A_{i} \lor \ldots \lor A_{j}\right)\right] \land \left(A \lor D\right)$ \\
with the following requirements \\
$|\left(\lnot B_{1}, \ldots, \lnot B_{i - 1}, \lnot C_{1}, \ldots, \lnot C_{i}, A \lor D\right)| = |\left(1, \ldots, m\right)|$, \\
$|\left(\lnot B_{1}, \ldots, \lnot B_{i - 1}\right)| = |\left(1, \ldots, i - 1\right)|$, \\
$|\left(\lnot C_{1}, \ldots, \lnot C_{i}\right)| = |\left(i, \ldots, j\right)|$ \\
$|\left(\lnot B_{1}, \ldots, \lnot B_{i - 1}, \lnot C_{1}, \ldots, \lnot C_{i}\right)| = |\left(1, \ldots, j\right)|$, \\
$\left(1, \ldots, i - 1, i, \ldots, j, m\right) = \left(1, \ldots, m\right)$, \\
$\left(1, \ldots, j, m\right) = \left(1, \ldots, m\right)$, \\
ensuring that the number of \subexpr{operands} equals the corresponding number of \subexpr{parameters} for the various ranges of indices of $A$ (that is, for $A_{1}, \ldots, A_{m}$, $A_{1}, \ldots, A_{i - 1}$, $A_{i}, \ldots, A_{j}$, and $A_{1}, \ldots, A_{j}$), and that the tuple of indices of the equivalent alternative expansions equal the original $\left(1, \ldots, m\right)$.  This example is also interesting because the internal quantification over $A_{i}, \ldots, A_{j}$ is masking a portion of the $\left(A_{1}, \ldots, A_{m}\right)$ that is being mapped.  Such masking is not advisable in practice, but it is an interesting test case.

\subsection{Automatic relabeling (capture avoidance)}
\label{Appendix:CaptureAvoidance}
When replacements (substitutions) are made, we must ensure that the meaning of \exprtype{Lambda} expressions are not altered by capturing variables which should be free.  In lambda calculus terminology, this is known as capture avoidance.  For example, given the expression \\
$a \mapsto a + b$ \\
we cannot replace $b$ with $f(a)$ unless we relabel the external $a$.  The original expression represents a function that accepts a single argument and yields that argument plus something that is free.  Replacing the free variable of this expression with something is not entirely free would change this meaning.

When such an invalid replacement is attempted in \ProveIt{}, rather than giving an error, we simply relabel the conflicting bound variable to an available dummy variable that does not have a conflict.  This is very much like the manner in which we choose dummy variables for making canonical forms of lambda expressions described in Appendix~\ref{Sec:Relabeling}.  
For the example above, replacing $b$ with $f(a)$ would produce \\
$_{-}a \mapsto _{-}a + f(a)$ \\
where $a$ and $f$ are free but $_{-}a$ is bound.
If there is a conflict and the variable is non-relabel-able for reasons discussed in Appendix~\ref{Sec:Relabeling}, there will be an error.

\subsection{Operation replacement}

\subsubsection{Explicit operation replacement} 

When an operator is replaced with a \exprtype{Lambda} expression, this lambda mapping is immediately applied as described in Appendix~\ref{Sec:Application}.  For example, $P(x, y)$ with \\
        $P: [(a, b) \mapsto a+b]$ \\
        becomes $x + y$.
This is not only convenient but necessary.
Recall from Appendix~\ref{Sec:ExprTypes} that an \exprtype{Operation} restricts what is allowed as an \subexpr{operator}.  In particular, a \exprtype{Lambda} expression is not allowed as an \subexpr{operator}.  But by immediately applying the lambda map (performing beta reduction), we avoid this issue and, importantly, we avoid Curry's paradox [Sec.~\ref{Sec:CurrysParadox}].

The one other restriction is that the \subexpr{body} of the \exprtype{Lambda} that is replacing the operator \emph{cannot} be an \exprtype{ExprRange} type of expression.  This is important to protect arity of operations, and lengths of tuples more generally.
Without this restriction, you could generate a contradiction (there would be a paradox).  For example,  you can prove that \\
$\vdash \forall_f \left|(f(a), b, c)\right| = 3$ \\
That is, the length of the tuple $(f(a), b, c)$ is $3$ regardless of what $f$ may be.
But if you then instantiate $f$ with \\
$f : x \mapsto x_1, \ldots, x_4$ \\
where the right side of $\mapsto$ is an \exprtype{ExprRange}, you \emph{would} derive \\
$\vdash \left|(a_1, \dots, a_4, b, c)\right| = 3$ \\
But, $\left|(a_1, \dots, a_4, b, c)\right|$ should be equal to $6$.  As long as $f(a)$ is not an \exprtype{ExprRange}, though, we will not generate such a contradiction.  So by disallowing the \exprtype{Lambda} \subexpr{body} to be an \exprtype{ExprRange}, we avoid such a contradiction.

\subsubsection{Implicit operation replacement}

When an operator is replaced with a designated \exprtype{Literal} operator of an \exprtype{Operation} class, the operation is replaced by an instance of that derived class rather than the \exprtype{Operation} class. For example, $P(x, y)$ with $P:+$ becomes $x + y$ with the correct \verb#Add# type rather than a generic \exprtype{Operation} type.  This is not an important reduction rule for an independent proof checker which is only concerned about primitive expression types, but it is important for the proper behavior of \ProveIt{} when generating a proof and formatting the resulting expressions.

\subsection{ExprRange expansion}
\label{sec:range_expansion}

There are two kinds of reduction rules that are specific to \exprtype{ExprRange} expressions that are applied in the process of performing a lambda application (beta reduction).

\subsubsection{Indexed variable expansions}

When the \exprtype{ExprRange} contains one or more \exprtype{IndexedVar} expressions with an index that is parameterized over the range, and the range of indexed variables is being replaced by an \exprtype{ExprRange}, it may be reduced via either
    \begin{description}
    \item[Parameter dependent expansion] is required whenever the parameter of the \exprtype{ExprRange} occurs anywhere besides as an index of an indexed variable being expanded.  In this case, the indices of the expanded indexed variable must match the original indices.  For example, \\
    $1 \cdot x_1 + \ldots + n \cdot x_n$ would be expanded, under \\
    $(x_1, \ldots, x_n) : (a_1, \ldots, a_j, a_{j+1}, \ldots, a_n)$ \\
    to \\
    $1 \cdot a_1 + \ldots + j \cdot a_j + (j+1) \cdot a_{j+1} + \ldots + n \cdot a_n$ \\
    assuming $0 \leq j \leq n$ and requiring \\
    $(1, \ldots, j, j+1, \ldots n) = (1, \ldots, n)$. \\  
    However, an error would occur given\\
    $(x_1, \ldots, x_n) : (a_1, \ldots, a_j, b_1, \ldots, b_k)$ \\
    as the expansion.
    \item[Parameter independent expansion] is allowed otherwise.  In this case, only the lengths of the expansion must match, not the indices themselves.  For example, $x_1 \cdot y_1 + \ldots + x_n \cdot y_n$ could be expanded, under\\
    $(x_1, \ldots, x_n) : (a_1, \ldots, a_j, b_1, \ldots, b_k)$ \\
    $(y_1, \ldots, y_n) : (c_1, \ldots, c_j, d_1, \ldots, d_k)$ \\
    to \\
    $a_1 \cdot c_1 + \ldots + a_j \cdot c_j + b_1 \cdot d_1 + \ldots + b_k \cdot d_k$ \\
    assuming $j + k = n$.
    \end{description}

\subsubsection{Range reductions}

When the \exprtype{ExprRange} is known to be empty or contain only a single element, it may be reduced via either
  \begin{description}
  \item[Empty range reduction] when the \exprtype{ExprRange} is known to be empty it will be replaced with zero elements 
  in a containing \exprtype{ExprTuple}.  For example, $(a, b_1, \ldots, b_n, c)$ with $n : 0$ reduces to $(a, c).$
  \item[Singular element range reduction] when the \exprtype{ExprRange} is known to contain only a single element it 
  will be replaced with this one element.  For example, $(a, b_1, \ldots, b_n, c)$ with $n : 1$ reduces to $(a, b_1, c).$
  \end{description}

\subsection{Conditional assumptions}

The \exprtype{Conditional} primitive expression type has one special reduction rule.  Specifically, the condition of the 
\exprtype{Conditional} may be introduced as an internal assumption in the process of performing a lambda application (or instantiation).  Note that \exprtype{Conditional}s are used implicitly in the conditions of quantifiers.  For example, $\forall_{x~|~Q(x)} P(x)$ is internally represented as an operation with the $\forall$ operator applied to the lambda expression $x \mapsto \left\{P(x) \textrm{ if } Q(x)\right..$, where $\left\{P(x) \textrm{ if } Q(x)\right..$ is a \exprtype{Conditional} expression, defined to be $P(x)$ when $Q(x)$ equals "true" but otherwise undefined for any practical purpose.  Since the $P(x)$ value of the \exprtype{Conditional} is only relevant when $Q(x)$ equals "true", \ProveIt{} will add $Q(x)$ to the 
assumptions when making replacements of the $P(x)$ expression in the process of performing a lambda application (e.g., during an instantiation).

\subsection{Automated equality reductions}

Automated equality reduction is simply a way to replace an expression with another one that is provable equal to the original during an instantiation, typically to avoid notational inconveniences.  For example, consider instantiating \\
$\vdash \forall_{n \in \mathbb{N^+}} \left[ \forall_{x_1, \ldots x_n, y_1, \ldots, y_n~|~(x_1=y_1) \land \ldots \land (x_n=y_n)}~
\begin{array}{l}
(x_1, \ldots, x_n) = \\
(y_1, \ldots, y_n)
\end{array}
\right]$ \\
with $n : 1$.  Without the automated range reduction feature, you would end up with \\
$\vdash \forall_{x_1, y_1~|~\land(x_1=y_1)} (x_1) = (y_1)$ \\
where $\land(x_1=y_1)$ represents conjunction applied to a single operand $(x_1 = y_1)$.  However, because we use automated range reduction for occurrences of unary conjunction, we instead obtain \\
$\vdash \forall_{x_1, y_1~|~x_1=y_1} (x_1) = (y_1)$ \\
directly, but it does require \\
$\vdash \land(x_1=y_1) = (x_1 = y_1)$ \\
as a requirement to effect this automated equality reduction.

Automated equality reductions can be implemented for any expression class to perform reductions as desired.  We prefer to limit its use, however, to instances where it avoids notational inconveniences (such as $\land(x_1 = y_1)$ which is notationally awkward).  The danger of over-using this feature is that it can limit a user's control to derive the exact expressions they are trying to derive.  For example, they may want to delay a reduction for pedagogical purposes.  While the feature can be disabled for individual expression classes, over-use of the feature can still be inconvenient and confusing to a user.  We therefore advise caution regarding the use of this feature.

When automated equality reductions are performed in an instantiation step, the requirements for performing these reductions (which are necessarily equality statements with the reducible expression on the left and the reduced version on the right) will be marked as ``equality replacement requirements.''  This is useful information for an independent proof checker (automated or manual) which simply needs to replace reducible expressions with their reduced versions while effecting the instantiation step to confirm that it was implemented faithfully.

%% file: axioms.tex
In this section, we list some of the basic theory packages of \ProveIt{} and their axioms.  Ideally, axioms provide the most fundamental definitions of the various kinds of useful expressions that may be constructed.

Each axiom is a self-evident or asserted truth, tied to a particular theory package, with a meaning that is governed by its expression DAG.  For simplicity, however, we only show a particular stylistic notation for each of these, with Table \ref{tab:symbols} as a legend of symbols for logic and set theory notation.  The full expression DAGs are available on the \ProveItWebsite.

These axioms are subject to change.  In some cases, the best choice for the axioms is not obvious.  They are chosen to be clear and fundamental, not necessarily minimalistic, and this is a subjective criteria.  

It is intended that these basic axioms be consistent with standard, well-established mathematical norms.  However, because we do not use intrinsic type restrictions or presume implicit domains of discourse, as discussed in Sec. \ref{Sec:DomainsOfDiscourse}, some of the axioms extend definitions beyond typical uses.  For example, rather than intrinsically limiting logical negation to Boolean types, we only provide definitions for the Boolean types and furthermore assert that if a logical negation has a Boolean value then its operand must be a Boolean value (see axiom 4 of the \texttt{proveit.logic.boolean.negation} axioms below).  This can allow us to conveniently infer the type of an expression indirectly.  We simply take an agnostic position with respect to the logical negation of a non-Boolean value; it is what it is and cannot be reduced.  We have similar axioms for logical conjunction and disjunction that allow us to conveniently infer the type of their operands if we know that the conjunction/disjunction is a Boolean.  Note, however, that we do not and can not, in our framework, similarly infer the type of an implication operand to be a Boolean.  An implication is always Boolean in \ProveIt{} regardless of the types of its operands.  One should regard $A \Rightarrow B$ in \ProveIt{} as no different from $(A=\top) \Rightarrow (B=\top)$, which is clearly either true or false.  This is not a consequence of its axioms (the \texttt{proveit.logic.boolean.implication} axioms all have Boolean type restrictions).  It is a consequence at a deeper core level of the lack of intrinsic type restrictions, the law of Deduction [see Appendix \ref{Sec:Deduction}], as well axiom 4 of \texttt{proveit.logic.booleans} axioms which allows us to deduce $\vdash A = \top$ from $\vdash A$.

\begin{table}
\begin{tabular}{c|c}
\hline
$\top$ & true  \\
\hline
$\bot$ & false \\
\hline
$\mathbb{B}$ & set of Boolean \\
& values (true or false) \\
\hline
$\Rightarrow$ & implication \\
\hline
$\lnot$ & logical negation (not) \\
\hline
$\land$ & logical conjunction (and) \\
\hline
$\lor$ & logical disjunction (or) \\
\hline
$\forall$ & universal quantification (forall) \\
\hline
$\exists$ & existential quantification (exists) \\
\hline
$\nexists$ & non-existence \\
\hline
$=$ & equals \\
\hline
$\neq$ & not equals \\
\hline
$\in$ & set membership \\
\hline
$\notin$ & set non-membership \\
\hline
$\cong$ & set equivalence \\
\hline
$\ncong$ & set non-equivalence \\
\hline
$\subset$ & proper (strict) subset \\
\hline
$\subseteq$ & subset \\
\hline
$\nsubset$ & not proper subset \\
\hline
$\nsubseteq$ & non-subset \\
\hline
$\supset$ & proper (strict) superset \\
\hline
$\supseteq$ & superset \\
\hline
$\nsupset$ & not proper superset \\
\hline
$\nsupseteq$ & non-superset \\
\hline
$\cup$ & set union \\
\hline
$\cap$ & set intersection \\
\hline
$\mathbb{N}$ & natural numbers \\
 & (whole number starting from zero) \\
\hline
$\mathbb{N}^{+}$ & natural numbers excluding zero \\
\hline
$\mathbb{Z}^{-}$, $\mathbb{Z}^{\neq 0}$, $\mathbb{Z}$& 
integers: negative, non-zero, and all \\
& whole numbers respectively \\
\hline
$\mathbb{Q}$ & rational numbers \\
\hline
$\mathbb{Q}^{\neq 0}$, $\mathbb{Q}^{+}$, & non-zero, positive, non-negative, and negative \\
$\mathbb{Q}^{\geq 0}$, $\mathbb{Q}^{-}$ & rational numbers respectively \\
\hline
$\mathbb{R}$ & real numbers \\
\hline
$\mathbb{R}^{\neq 0}$, $\mathbb{R}^{+}$, & non-zero, positive, non-negative, and negative \\
$\mathbb{R}^{\geq 0}$, $\mathbb{R}^{-}$ & real numbers respectively \\
\hline
$\mathbb{C}$ & complex numbers \\
\hline
\end{tabular}
\caption{Legend of symbols for logic and set theory notation and number sets.} 
\label{tab:symbols}
\end{table}

\begin{itemize}

\item \texttt{proveit.core\_expr\_types.operations} axiom
    \begin{enumerate}
    \item $\vdash ~\left[
    \begin{array}{l}
    \forall_{n \in \mathbb{N}} \forall_{f, x_{1}, \ldots, x_{n}, y_{1}, \ldots, y_{n} ~|~\left(x_{1}, \ldots, x_{n}\right) = \left(y_{1}, \ldots, y_{n}\right)} \\
    \left(f(x_{1}, \ldots, x_{n}) = f(y_{1}, \ldots, y_{n})\right)
    \end{array}
    \right]$
    \end{enumerate}

\item \texttt{proveit.core\_expr\_types.conditionals} axioms
    \begin{enumerate}
    \item $\vdash \forall_{a}~\left(\left\{a \textrm{ if } \top\right.. = a\right)$ 
    \item $\vdash \forall_{a, Q}~\left(\begin{array}{c} \left\{a \textrm{ if } Q\right.. \\  = \left\{a \textrm{ if } Q = \top\right.. \end{array}\right)$
    \item $\vdash \forall_{a, b, Q~|~Q \Rightarrow \left(a = b\right)}~\left(\begin{array}{c} \left\{a \textrm{ if } Q\right.. \\  = \left\{b \textrm{ if } Q\right.. \end{array}\right)$
    \end{enumerate}
    
\item \texttt{proveit.core\_expr\_types.lambda\_maps} axiom
    \begin{enumerate}
    \item $\vdash \left[
    \begin{array}{l}
    \forall_{i \in \mathbb{N}^+}~\forall_{f, g} \\
    \left(\begin{array}{c} 
    \left[\forall_{a_{1}, \ldots, a_{i}}~\left(
    \begin{array}{c}
    f(a_{1}, \ldots, a_{i}) = \\
    g(a_{1}, \ldots, a_{i})
    \end{array}
    \right)\right] \Rightarrow  \\ \left(\begin{array}{c} \left[\left(b_{1}, \ldots, b_{i}\right) \mapsto f(b_{1}, \ldots, b_{i})\right] =  \\ \left[\left(c_{1}, \ldots, c_{i}\right) \mapsto g(c_{1}, \ldots, c_{i})\right] \end{array}\right) \end{array}\right)
    \end{array}
    \right]$
    \end{enumerate}

\item \texttt{proveit.core\_expr\_types.tuples} axioms
    \begin{enumerate}
    \item $\vdash |()| = 0$
    \item $\vdash \forall_{i \in \mathbb{N}}~\left[\forall_{a_{1}, \ldots, a_{i}, b}~\left(|\left(a_{1}, \ldots, a_{i}, b\right)| = \left(i + 1\right)\right)\right]$
    \item $\vdash 
    \left[\begin{array}{l}
    \forall_{i \in \mathbb{N}}~\forall_{a_{1}, \ldots, a_{i}, b, c_{1}, \ldots, c_{i}, d} \\
    \left(
    \begin{array}{c} 
    \left(
    \begin{array}{c}
    \left(a_{1}, \ldots, a_{i}, b\right) = \\
    \left(c_{1}, \ldots, c_{i}, d\right)
    \end{array}
    \right) =  \\ 
    \left( \left(
    \begin{array}{c}
    \left(a_{1}, \ldots, a_{i}\right) = \\
    \left(c_{1}, \ldots, c_{i}\right)
    \end{array} 
    \right) \land \left(b = d\right)
    \right)
    \end{array}\right)
    \end{array}\right]
    $
    \item $\vdash \forall_{f, i, j~|~\left(j + 1\right) = i}~\left(\left(f(i), \ldots, f(j)\right) = ()\right)$
    \item $\vdash \forall_{f, i, j~|~|\left(f(i), \ldots, f(j)\right)| \in \mathbb{N}}~\left(\begin{array}{c} \left(f(i), \ldots, f(j + 1)\right) =  \\ \left(f(i), \ldots, f(j), f(j + 1)\right) \end{array}\right)$
    \end{enumerate}

\item \texttt{proveit.logic.booleans} axioms
    \begin{enumerate}
    \item $\vdash \top$
    \item $\vdash \mathbb{B} = \{\top, \bot\}$ 
    \item $\vdash \bot \neq \top$
    \item $\vdash \forall_{A~|~A}~\left(A = \top\right)$ 
    \item $\vdash \forall_{A~|~A = \top}~A$
    \end{enumerate}

\item \texttt{proveit.logic.booleans.implication} axioms
    \begin{enumerate}
    \item $\vdash \left(\top \Rightarrow \bot\right) = \bot$
    \item $\vdash \forall_{A \in \mathbb{B}~|~(\lnot A) \Rightarrow \bot}~A$
    \item $\vdash \forall_{A \in \mathbb{B}~|~A \Rightarrow \bot}~(\lnot A)$
    \item $\vdash \forall_{A, B}~\left(\left(A \Leftrightarrow B\right) = \left(\left(A \Rightarrow B\right) \land \left(B \Rightarrow A\right)\right)\right)$
    \end{enumerate}

\item \texttt{proveit.logic.booleans.negation} axioms 
    \begin{enumerate}
    \item $\vdash \lnot \top = \bot$
    \item $\vdash \lnot \bot = \top$
    \item $\vdash \forall_{A~|~\lnot A}~\left(A = \bot\right)$
    \item $\vdash \forall_{A~|~\lnot A \in \mathbb{B}}~A \in \mathbb{B}$
    \end{enumerate}

\item \texttt{proveit.logic.booleans.conjunction} axioms
    \begin{enumerate}
    \item $\vdash \left(\top \land \top\right) = \top$
    \item $\vdash \left(\top \land \bot\right) = \bot$
    \item $\vdash \left(\bot \land \top\right) = \bot$ 
    \item $\vdash \left(\bot \land \bot\right) = \bot$ 
    \item $\vdash \forall_{A, B~|~\left(A \land B\right) \in \mathbb{B}}~\left(A \in \mathbb{B}\right)$
    \item $\vdash \forall_{A, B~|~\left(A \land B\right) \in \mathbb{B}}~\left(B \in \mathbb{B}\right)$
    \item $\vdash \left[\land\right]\left(\right)$ 
    \item $\vdash \forall_{m \in \mathbb{N}}~\left[\forall_{A_{1}, \ldots, A_{m}, B}~\left(\begin{array}{c} \left(A_{1} \land \ldots \land A_{m} \land B\right) =  \\ \left(\left(A_{1} \land \ldots \land A_{m}\right) \land B\right) \end{array}\right)\right]$
    \end{enumerate}

\item \texttt{proveit.logic.booleans.disjunction} axioms
    \begin{enumerate}
    \item $\vdash \left(\top \lor \top\right) = \top$
    \item $\vdash \left(\top \lor \bot\right) = \top$
    \item $\vdash \left(\bot \lor \top\right) = \top$
    \item $\vdash \left(\bot \lor \bot\right) = \bot$
    \item $\vdash \forall_{A, B~|~\left(A \lor B\right) \in \mathbb{B}}~\left(A \in \mathbb{B}\right)$ 
    \item $\vdash \forall_{A, B~|~\left(A \lor B\right) \in \mathbb{B}}~\left(B \in \mathbb{B}\right)$
    \item $\vdash \lnot \left[\lor\right]\left(\right)$
    \item $\vdash \forall_{m \in \mathbb{N}}~\left[\forall_{A_{1}, \ldots, A_{m}, B}~\left(\begin{array}{c} \left(A_{1} \lor \ldots \lor A_{m} \lor B\right) =  \\ \left(\left(A_{1} \lor \ldots \lor A_{m}\right) \lor B\right) \end{array}\right)\right]$
    \end{enumerate}

\item \texttt{proveit.logic.booleans.quantification.universality} axiom
    \begin{enumerate}
    \item $\vdash \forall_{n \in \mathbb{N}^+}~\left[\forall_{P}~\left(\left[\forall_{x_{1}, \ldots, x_{n}}~P(x_{1}, \ldots, x_{n})\right] \in \mathbb{B}\right)\right]$
    \end{enumerate}

\item \texttt{proveit.logic.booleans.quantification.existence} axioms
    \begin{enumerate}
    \item $\vdash \forall_{n \in \mathbb{N}^+}~\left[\forall_{P, Q}~\left(\begin{array}{c} \left[\exists_{x_{1}, \ldots, x_{n}~|~Q(x_{1}, \ldots, x_{n})}~P(x_{1}, \ldots, x_{n})\right] =  \\ 
    \lnot 
    \left[
    \begin{array}{l}
    \forall_{y_{1}, \ldots, y_{n}~|~Q(y_{1}, \ldots, y_{n})}\\
    \left(P(y_{1}, \ldots, y_{n}) \neq \top\right)
    \end{array}
    \right] \end{array}\right)\right]$
    \item $\vdash \forall_{n \in \mathbb{N}^+}~\left[\forall_{P, Q}~\left(\begin{array}{c} \left[\nexists_{x_{1}, \ldots, x_{n}~|~Q(x_{1}, \ldots, x_{n})}~P(x_{1}, \ldots, x_{n})\right] =  \\ \lnot \left[\exists_{y_{1}, \ldots, y_{n}~|~Q(y_{1}, \ldots, y_{n})}~P(y_{1}, \ldots, y_{n})\right] \end{array}\right)\right]$
    \end{enumerate}

\item \texttt{proveit.logic.equality} axioms
    \begin{enumerate}
    \item $\vdash \forall_{x, y}~\left(\left(x = y\right) \in \mathbb{B}\right)$
    \item $\vdash \forall_{x, y}~\left(\left(y = x\right) = \left(x = y\right)\right)$
    \item $\vdash \forall_{x, y, z~|~x = y, y = z}~\left(x = z\right)$ 
    \item $\vdash \forall_{x, y}~\left(\left(x \neq y\right) = (\lnot \left(x = y\right))\right)$
    \item $\vdash \forall_{x}~\left(x = x\right)$
    \item $\vdash \forall_{f, x, y~|~x = y}~\left(f(x) = f(y)\right)$
    \end{enumerate}

\item \texttt{proveit.logic.sets.membership} axioms
    \begin{enumerate}
    \item $\vdash \forall_{x, S}~\left(\left(x \notin S\right) = (\lnot \left(x \in S\right))\right)$
    \end{enumerate}
    
\item \texttt{proveit.logic.sets.equivalence} axiom
    \begin{enumerate}
        \item $\vdash \forall_{A, B}~\left(\left(A \cong B\right) = \left[\forall_{x}~\left(\left(x \in A\right) = \left(x \in B\right)\right)\right]\right)$
        \item $\vdash \forall_{A, B}~\left(\left(A \ncong B\right) = (\lnot \left(A \cong B\right))\right)$
    \end{enumerate}

\item \texttt{proveit.logic.sets.enumeration} axiom
    \begin{enumerate}
    \item $\vdash \forall_{n \in \mathbb{N}}~\left[\forall_{x, y_{1}, \ldots, y_{n}}~\left(\begin{array}{c} \left(x \in \left\{y_{1}, \ldots, y_{n}\right\}\right) =  \\ \left(\left(x = y_{1}\right) \lor \ldots \lor \left(x = y_{n}\right)\right) \end{array}\right)\right]$
    \end{enumerate}

\item \texttt{proveit.logic.sets.inclusion} axioms
    \begin{enumerate}
        \item $\vdash \forall_{A, B}~\left(\left(A \subseteq B\right) = \left[\forall_{x \in A}~\left(x \in B\right)\right]\right)$
        \item $\vdash \forall_{A, B}~\left(\left(A \nsubseteq B\right) = (\lnot \left(A \subseteq B\right))\right)$
        \item $\vdash \forall_{A, B}~\left(\left(A \subset B\right) = \left(\left(A \subseteq B\right) \land \left(B \ncong A\right)\right)\right)$
        \item $\vdash \forall_{A, B}~\left(\left(A \not\subset B\right) = (\lnot \left(A \subset B\right))\right)$
    \end{enumerate}

\item \texttt{proveit.logic.sets.unification} axioms
    \begin{enumerate}
    \item $\vdash \forall_{m \in \mathbb{N}^+}~\left[\forall_{x, A_{1}, \ldots, A_{m}}~\left(\begin{array}{c} \left(x \in \left(A_{1} \cup \ldots \cup A_{m}\right)\right) =  \\ \left(\left(x \in A_{1}\right) \lor \ldots \lor \left(x \in A_{m}\right)\right) \end{array}\right)\right]$
    \item $\vdash 
    \left[
    \begin{array}{l}
    \forall_{n \in \mathbb{N}^+} \forall_{S_{1}, \ldots, S_{n}, Q, R, x}~\\
    \left(\begin{array}{c} \left(
    x \in \left[\bigcup\limits_{y_{1} \in S_{1}, \ldots, y_{n} \in S_{n}~|~Q(y_{1}, \ldots, y_{n})}~R(y_{1}, \ldots, y_{n})\right]\right) \\  = \left[
    \begin{array}{l}
    \exists_{y_{1} \in S_{1}, \ldots, y_{n} \in S_{n}~|~Q(y_{1}, \ldots, y_{n})}~\\ \left(x \in R(y_{1}, \ldots, y_{n})\right)
    \end{array}
    \right] \end{array}\right)
    \end{array}
    \right]$
    \end{enumerate}

\item \texttt{proveit.logic.sets.intersection} axioms
    \begin{enumerate}
    \item $\vdash \forall_{m \in \mathbb{N}^+}~\left[\forall_{x, A_{1}, \ldots, A_{m}}~\left(\begin{array}{c} \left(x \in \left(A_{1} \cap \ldots \cap A_{m}\right)\right) =  \\ \left(\left(x \in A_{1}\right) \land \ldots \land \left(x \in A_{m}\right)\right) \end{array}\right)\right]$
    \item $\vdash 
    \left[
    \begin{array}{l}
    \forall_{n \in \mathbb{N}^+} \forall_{S_{1}, \ldots, S_{n}, Q, R, x~|~\exists_{y_{1} \in S_{1}, \ldots, y_{n} \in S_{n}}~Q(y_{1}, \ldots, y_{n})}~\\
    \left(\begin{array}{c} \left(x \in \left[\bigcap\limits_{y_{1} \in S_{1}, \ldots, y_{n} \in S_{n}~|~Q(y_{1}, \ldots, y_{n})}~R(y_{1}, \ldots, y_{n})\right]\right) \\  = \left[
    \begin{array}{l}
    \forall_{y_{1} \in S_{1}, \ldots, y_{n} \in S_{n}~|~Q(y_{1}, \ldots, y_{n})}~\\
    \left(x \in R(y_{1}, \ldots, y_{n})\right)
    \end{array}
    \right] 
    \end{array}\right)
    \end{array}
    \right]$
    \end{enumerate}

\item \texttt{proveit.logic.sets.subtraction} axiom
    \begin{enumerate}
    \item $\vdash \forall_{x, A, B}~\left(\left(x \in \left(A - B\right)\right) = \left(\left(x \in A\right) \land \left(x \notin B\right)\right)\right)$
    \end{enumerate}

\item \texttt{proveit.logic.sets.comprehension} axiom
    \begin{enumerate}
    \item $\vdash
    \left[
    \begin{array}{l}
    \forall_{n \in \mathbb{N}^+} \forall_{S_{1}, \ldots, S_{n}, Q, f, x}~\\
    \left(
    \begin{array}{c} \left(x \in 
    \left\{
    \begin{array}{c}
    f(y_{1}, \ldots, y_{n})~:~ \\
    Q(y_{1}, \ldots, y_{n})
    \end{array}
    \right\}_{y_{1} \in S_{1}, \ldots, y_{n} \in S_{n}}\right) \\  = 
    \left[
    \begin{array}{l} \exists_{y_{1} \in S_{1}, \ldots, y_{n} \in S_{n}~|~Q(y_{1}, \ldots, y_{n})} \\
    \left(x = f(y_{1}, \ldots, y_{n})\right)
    \end{array}\right] 
    \end{array}\right)
    \end{array}\right]$
    \end{enumerate}

\item \texttt{proveit.logic.sets.power\_sets} axiom
    \begin{enumerate}
        \item $\vdash \forall_{x, S}~\left(\left(x \in \mathbb{P}(S)\right) = \left(x \subseteq S\right)\right)$
    \end{enumerate}

\item \texttt{proveit.logic.sets.cardinality} axioms
    \begin{enumerate}
    \item $\vdash |\emptyset| = 0$
    \item $\vdash \forall_{x, S~|~|S| \in \mathbb{N}, x \notin S}~\left(|S \cup \left\{x\right\}| = \left(|S| + 1\right)\right)$ 
    \item Cardinality of infinite sets are not yet defined (requires ordinal numbers and additional axioms).
    \end{enumerate}

\item \texttt{proveit.numbers.number\_sets.natural\_numbers} axioms (1-5 are Peano's axioms)
    \begin{enumerate}
    \item $\vdash 0 \in \mathbb{N}$ 
    \item $\vdash \forall_{n \in \mathbb{N}}~\left(\left(n + 1\right) \in \mathbb{N}\right)$ 
    \item $\vdash \forall_{m \in \mathbb{N}, n \in \mathbb{N}~|~\left(m + 1\right) = \left(n + 1\right)}~\left(n = m\right)$ 
    \item $\vdash \forall_{n \in \mathbb{N}}~\left(\left(n + 1\right) \neq 0\right)$ 
    \item $\vdash \forall_{S~|~S \subseteq \mathbb{N}}~\left(\left(\left(0 \in S\right) \land \left[\forall_{x \in S}~\left(\left(x + 1\right) \in S\right)\right]\right) \Rightarrow \left(S \cong \mathbb{N}\right)\right)$
    \item $\vdash \forall_{x}~\left(\left(x \in \mathbb{N}\right) \in \mathbb{B}\right)$
    \item $\vdash \mathbb{N}^+ = \left\{n~|~n > 0\right\}_{n \in \mathbb{N}}$
    \end{enumerate}

\end{itemize}